\documentclass[twocolumn,twoside]{IEEEtran}
\usepackage{mathpple}
\usepackage{times}
\usepackage{amsmath}  
\usepackage{amssymb}  
\usepackage{mathrsfs} 
\usepackage{theorem}  
\usepackage{comment}  
\usepackage{upref}
\usepackage{amsfonts}
\usepackage{verbatim}
\usepackage[dvipsnames,usenames]{color}
\usepackage{enumerate}
\usepackage{graphicx}
\usepackage{subfigure}
\usepackage{latexsym}
\usepackage{mathtools} 
\usepackage{array, tabularx} 
\usepackage[shortlabels]{enumitem}
\usepackage{url}
\usepackage{bbm}

\usepackage{times,color}
\usepackage{enumitem}
\usepackage{float}
\usepackage{listings}
\usepackage{tikz}
\usepackage{multicol}
\usepackage{psfrag}
\usepackage{pgfplots}
\usepackage{hhline}
\usepackage{xcolor,colortbl}
\usepackage{algorithm,algpseudocode,refcount} 

\renewcommand{\algorithmicrequire}{\textbf{Input:}}
\renewcommand{\algorithmicensure}{\textbf{Output:}}
\algnewcommand\algorithmiccase{\textbf{case}}
\algdef{SE}[CASE]{Case}{EndCase}[1]{\algorithmiccase\ #1}{\algorithmicend\ \algorithmiccase}%
\algtext*{EndCase}%
\makeatletter
\newcommand{\removelatexerror}{\let\@latex@error\@gobble}

\newcommand{\plus}{\scalebox{0.75}[1.0]{$+$}}

\newcommand{\limup}[1]{\lim_{#1\rightarrow\infty}}

\newcommand{\0}{\mathbf{0}}

\newcommand{\be}[1]{\begin{equation}\label{#1}}
\newcommand{\ee}{\end{equation}}

\newcommand{\bc}{\begin{center}}
	\newcommand{\ec}{\end{center}}

\newcommand{\floorenv}[1]{\left\lfloor #1 \right\rfloor}

\newcommand{\ceilenv}[1]{\left\lceil #1 \right\rceil}

\newcommand{\cB}{{\cal B}}
\newcommand{\cC}{{\cal C}}
\newcommand{\cD}{{\cal D}}

\newcommand{\cO}{{\cal O}}

\newcommand{\cR}{{\cal R}}

\newcommand{\cZ}{{\cal Z}}

\newcommand{\bfs}{{\boldsymbol s}}

\newcommand{\bfu}{{\boldsymbol u}}
\newcommand{\bfv}{{\boldsymbol v}}
\newcommand{\bfw}{{\boldsymbol w}}
\newcommand{\bfx}{{\boldsymbol x}}
\newcommand{\bfy}{{\boldsymbol y}}
\newcommand{\bfz}{{\boldsymbol z}}

\newcommand{\bfW}{{\mathbf W}}

\renewcommand{\le}{\leqslant}
\renewcommand{\leq}{\leqslant}
\renewcommand{\ge}{\geqslant}
\renewcommand{\geq}{\geqslant}

\newcommand{\Cref}[1]{Co\-rol\-la\-ry\,\ref{#1}}

\theoremstyle{plain} \theorembodyfont{\normalfont\slshape}

\newtheorem{thm}{Theorem$\!$}
\newenvironment{theorem}{\begin{thm}\hspace*{-1ex}{\bf.}}{\end{thm}}

\newtheorem{prop}[thm]{Proposition$\!$}

\newtheorem{lem}[thm]{Lemma$\!$}
\newenvironment{lemma}{\begin{lem}\hspace*{-1ex}{\bf.}}{\end{lem}}

\newtheorem{cor}[thm]{Corollary$\!$}

\newtheorem{cons}[thm]{Construction$\!$}

\newtheorem{defi}[thm]{Definition$\!$}
\newenvironment{definition}{\begin{defi}\hspace*{-1ex}{\bf.}}{\end{defi}}

\newtheorem{cl}[thm]{Claim}
\newenvironment{claim}{\begin{cl}\hspace*{-1ex}{\bf .}}{\end{cl}}
\theorembodyfont{\normalfont}

\newtheorem{exam}{Example$\!$}
\newenvironment{example}{\begin{exam}\hspace*{-1ex}{\bf .}}{\end{exam}}

\newtheorem{remrk}{Remark$\!$}
\newenvironment{remark}{\begin{remrk}\hspace*{-1ex}{\bf .}}{\end{remrk}}

\definecolor{Codecolor}{named}{White}  

\newcommand{\Copen}{\mbox{\{\kern-5.50pt\{}}
\newcommand{\Cclose}{\mbox{\}\kern-5.50pt\}}}
\newcommand{\Cslash}{\mbox{$\backslash\kern-6.02pt\backslash$}}

\newcommand{\SubDistSet}{{\cZ}}
\newcommand{\SubDistSize}{{Z}}

\begin{document}
	\title{Reconstruction of Strings from their Substrings Spectrum}
	
	\author{\large Sagi~Marcovich,~\IEEEmembership{Student Member,~IEEE} and Eitan~Yaakobi,~\IEEEmembership{Senior Member,~IEEE} 
	\thanks{This work was presented in part at the IEEE International Symposium on Information Theory (ISIT), Los Angeles, CA, June 2020 (reference~\cite{MY20}).}
		\thanks{S. Marcovich and E. Yaakobi are with the Department of Computer Science, Technion --- Israel Institute of Technology, Haifa 3200003, Israel (e-mail: \texttt{\{sagimar,yaakobi\}@cs.technion.ac.il}).}
	}

	\maketitle
	\begin{abstract}
	This paper studies reconstruction of strings based upon their substrings spectrum. Under this paradigm, it is assumed that all substrings of some fixed length are received and the goal is to reconstruct the string. While many existing works assumed that substrings are received error free, we follow in this paper the noisy setup of this problem that was first studied by Gabrys and Milenkovic. The goal of this study is twofold. First we study the setup in which not all substrings in the multispectrum are received, and then we focus on the case where the read substrings are not error free. In each case we provide specific code constructions of strings that their reconstruction is guaranteed even in the presence of failure in either model. We present efficient encoding and decoding maps and analyze the cardinality of the code constructions, while studying the cases where the rates of our codes approach 1.
	\end{abstract}

	\section{Introduction} \label{sec:intro}
	In many storage and communication channels it is not possible to receive the transmitted or stored string as one unit, even in its noisy version. Rather, the information about the string can only be provided in several other forms such as a list of its subsequences, statistics on its symbols, and more. This class of models usually falls under the general framework of the \emph{string reconstruction problems}. There are several instances of this setup, such as the \emph{$k$-deck problem}~\cite{DudSch2003,BenMeySchSmiSto91,Sco97} and the \emph{reconstruction from substring compositions problem}~\cite{AchDasMilOrlPan10,AchDasMilOrlPan15,MotBreTse13, MoRaTs13, GaMoRa16, BreBreTse13, SCT15, ShKaXiCoTs16}. Similar problems under this paradigm are the \emph{trace reconstruction problem}~\cite{BatKanKhaMcG2004} and the \emph{reconstruction problem} by Levenshtein~\cite{Lev01}, however in these setups the string is received as one unit multiple times with possible errors.


This paper studies an important setup for this class of problems, where it is assumed that the information about the string is conveyed by the multispectrum of all its substrings of some fixed length. Under this paradigm, the goal is to reconstruct the string and the success of this process usually depends on the length of the read substrings and the stored string. This model of strings reconstruction is motivated by current DNA sequencing technologies and in particular shotgun DNA sequencing~\cite{Li15}. In this method, the DNA strand is broken into multiple fragments, called \emph{reads}, which are then assembled together to reconstruct the strand~\cite{Chin_2013,LoNiQuJoSiJa15,salzberg_2010}. 

Mathematically speaking, for a length-$n$ string $\bfw$ and a positive integer $L$, its \emph{$L$-multispectrum}, denoted by $S_L(\bfw)$, is the multiset of all its length-$L$ substrings, $S_L(\bfw) = \{\bfw_{1,L}, \bfw_{2,L},\ldots, \bfw_{n-L+1,L}\}$, where $\bfw_{i,L}$ is the substring $(w_i,w_{i+1},\ldots,w_{i+L-1})$. Then, the goal is to reconstruct the string $\bfw$ given its multispectrum $S_L(\bfw)$. If a string can be uniquely reconstructed from its $L$-multispectrum, then it is called \emph{$L$-reconstructible}. It was proved by Ukkonen~\cite{Ukk1992} that if all length-$(L-1)$ substrings of $\bfw$ are different from each other, then the string $\bfw$ is $L$-reconstructible. A string $\bfw$ that satisfies this constraint is referred as \emph{$(L-1)$-substring unique}. Based upon this property, it was recently proved in~\cite{GabMil18,EliGabMedYaa19} that if $L=\lceil a\log(n)\rceil$ for some fixed value of $a>1$, then the asymptotic rate of all $L$-reconstructible strings approaches 1. 

Several recent papers have taken an information-theoretic point of view to the string assembly problem. The goal of these works was to study the fundamental limits of reconstructing $ \bfw $ from $ S_L(\bfw) $ with a fixed failure probability under different setups and various error models. Arratia et al.~\cite{ArMaReWa96} studied the limits of any assembly algorithm that recovers $ \bfw $ from $ S_L(\bfw) $ where $ \bfw $ is an i.i.d DNA string, and later Motahari, Bresler, and Tse~\cite{MotBreTse13} studied the case where only a subset of $ S_L(\bfw) $ is available, while each read begins at a uniformly distributed location of the string. They both showed that if the reads are long enough to have no repeats, then reconstruction is possible with high probability.
This was then extended in~\cite{MoRaTs13} for the case where every read is transferred through a symmetric substitution noisy read channel 
and in~\cite{GaMoRa16} it was assumed that the reads are corrupted by at most some fixed number of edit errors. 
Moreover, it has been shown that if $ \bfw $ satisfies several constraints, which are based on its repeats statistics, then it can be assembled with high probability from $ S_L(\bfw) $~\cite{BreBreTse13} or a subset of it~\cite{SCT15}. 
Another variation of the string assembly problem, which allows to partially reconstruct the string $ \bfw $, was studied in~\cite{ShKaXiCoTs16}. 

In this paper, we follow the recent works by Gabrys and Milenkovic~\cite{GabMil18} and by Chang et al.~\cite{ChaChrEzeKia17} and assume that the $L$-multispectrum is not received error free, while still requiring to reconstruct $ \bfw $ in the \emph{worst case}. We consider two models of this setup. In the first one, it is assumed that not all substrings in the $L$-multispectrum were read so only a subset of $S_L(\bfw)$ is received. The second model assumes that all reads in the $L$-multispectrum were received, however some of them might be erroneous. An important tool in our constructions uses the set of substring unique strings and we also study its extension. Namely, for fixed $L$ and $d$, it is said that $\bfw$ is an \emph{$(L,d)$-substring distant} string if the Hamming distance between any two of its length-$L$ substrings is at least $d$. We study the cardinality of this set of strings and show an encoding and decoding maps for this constraint. 

While the motivation of this problem originates from shotgun DNA sequencing where the fragments, i.e. substrings, of the DNA strands are read and should be used to reconstruct the DNA strand~\cite{Li15}, the coding part does not match this model since natural DNA can be arbitrary and we do not necessarily have control on its data. However, we find the coding part to be important for DNA storage systems~\cite{CGK12,GBCDLSB13,GHPPS15,BLCCSS16,YTMZM15} where it is possible to encode the data and especially for future ones where long DNA strands will be able to be synthesized. Lastly, we consider in the paper only substitution errors, which were observed to be dominant errors in several previous DNA storage experiments~\cite{solqc}, while the extension for edit errors is left for future work.

The rest of the paper is organized as follows. In Section~\ref{sec:def}, we formally define the codes and constraints studied in this paper and review several previous results. In Section~\ref{sec:rec-incomplete}, we study the case where an incomplete multispectrum is received. Section~\ref{sec:rec-Erroneous} studies the setup where some of the read substrings are noisy as well as $(L,d)$-substring distant strings. Another construction for noisy substrings is presented in Section~\ref{sec:rec-Erroneous2}. Finally, Section~\ref{sec:concl} concludes the paper.

	\section{Definitions and Preliminaries}\label{sec:def}
	
In this section we formally define the notations, codes, and constraints studied in the paper. 
For an integer $ i \in \mathbb{N} $ we denote by $ [i] $ the set $ \{1, \dots, i\} $. For a multiset $ A $, let $ |A| $ denote the number of elements in $A$ (with repetitions). For a set $I$ of integers and $ a \in I $, $ b_I(a) $ denotes the binary representation of the index of $ a $ in $ I $ using $ \ceilenv{\log (|I|)} $ bits, when the integers are ordered in an increasing order. When $ I $ is omitted, it is implied that $ I = [n] $, while $n$ will be clear from the context.

Let $ \Sigma = \{0,1\} $ denote the binary alphabet, $ n $ an integer, and $ \bfw \in \Sigma^n $ a string. For two positive integers $i$ and $k$ such that $ i+k-1 \le n $, let $ \bfw_{i,k} $ denote the length-$k$ substring of $ \bfw $ starting at position $ i $. Additionally, let Pref$_k(\bfw) = \bfw_{1,k}, $ Suff$_k(\bfw) = \bfw_{n + 1 - k, k} $ denote the \emph{$k$-prefix}, \emph{$k$-suffix} of $\bfw$, respectively. For two strings $\bfw,\bfx\in \Sigma^n$, $d_H(\bfw,\bfx)$ is the Hamming distance between $\bfw$ and $\bfx$ and $w_H(\bfw)  = d_H(\bfw, \0)$ is the Hamming weight of $\bfw$. For a multiset $S=\{\bfs_1,\ldots,\bfs_m\}\subseteq \Sigma^n$ of strings, $d_H(S)$ is defined to be the \emph{minimum Hamming distance  of $S$}, which is the minimum Hamming distance among all pairs of strings in $S$, i.e., $d_H(S) =\min_{1\leq i < j\leq m}\{d_H(\bfs_i,\bfs_j)\}$. For a nonnegative integer $ r \le n $, $ \cB_{r}(\bfw) $ denotes the radius-$r$ Hamming ball around $ \bfw $, that is, $ \cB_{r}(\bfw) = \{\bfx\in \Sigma^n \ | \ d_H(\bfw,\bfx)\leq r \}$.

For a string $\bfw\in\Sigma^n$ and a positive integer $L\leq n$, the set $S_L(\bfw)$ is defined to be the \emph{$L$-multispectrum of $ \bfw $}, which is the \emph{multiset} of all its length-$L$ substrings
\[S_L(\bfw) = \{\bfw_{1,L}, \bfw_{2,L}, \ldots , \bfw_{n-L+1,L}\}.\]
The main family of strings studied in this paper is defined in the next definition. 
\begin{definition} 
A string $\bfw\in\Sigma^n$ is called an \textbf{$(L,d)$-substring distant string} if the Hamming distance of its $L$-multispectrum is at least $d$, that is, $d_H(S_L(\bfw))\geq d$. For $d=1$, we refer to an $(L,1)$-substring distance string as an \textbf{$L$-substring unique string}. 
\end{definition}
We note that the case of $d=1$ has also been studied in~\cite{EliGabMedYaa19} and was referred as \emph{repeat-free words}. Clearly, for given $n$ and $d$, the number of $ (L,d) $-substring distant strings grows as $ L $ increases. This follows from the observation that if $ \bfw $ is $ (L,d) $-substring distant, then it is immediately derived that $ \bfw $ is also $ (L',d) $-substring distant for every $ L < L' \le n $.


\begin{example}\label{ex1}
Let $ n=16$ and let \[ \bfx = 0100000111011111, \]
so its $ 6 $-multispectrum is 
	\begin{align*}
		S_6(\bfx) = & \{010000, 100000, 000001, 000011,000111,& \\
		& 001110, 011101, 111011, 110111, 101111,011111\}. &
	\end{align*}
The string $ \bfx $ is $ (6,1) $-substring distant, however, $ \bfx $ is not $ (6,2) $-substring distant since 
$$d_H(\bfx_{7,6},\bfx_{11,6}) = d_H(011101,011111) = 1.$$
\end{example}

The family of $(L,d)$-substring distant strings and more specifically $L$-substring unique strings is highly related to the set of reconstructible strings, which is defined next. Namely, a string $\bfw\in\Sigma^n$ is called an \textbf{\emph{$L$-reconstructible string}} if it can be uniquely reconstructed from its $L$-multispectrum. Hence, $\bfw$ is an $L$-reconstructible string if for every $ \bfx \in \Sigma^n $ where $\bfx\neq \bfw$ it holds that $S_L(\bfw)\neq S_L(\bfx)$. For positive integers $n,d,L$, we denote by $\SubDistSet_{n}(L,d)$ the set of all length-$n$ $(L,d)$-substring distant strings over $\Sigma$. For $d=1$ we simply denote this value by $\SubDistSet_{n}(L)$. The set of $L$-reconstructible strings is denoted by $\cR_{n}(L)$. We also let $\SubDistSize_{n}(L,d) = |\SubDistSet_{n}(L,d)|$, $\SubDistSize_{n}(L) = |\SubDistSet_{n}(L)|$, and $R_{n}(L) = |\cR_{n}(L)|$. 

The following connection between substring unique and reconstructible strings was first established by Ukkonen in~\cite{Ukk1992}. 
\begin{theorem}{\cite{Ukk1992}}\label{th:Ukk}
If a string $\bfx\in\Sigma^n$ is $(L-1)$-substring unique then it is $L$-reconstructible.
\end{theorem}
According to Theorem~\ref{th:Ukk}, it holds that $\SubDistSet_{n}(L-1) \subseteq \cR_{n}(L)$ and in particular $\SubDistSize_{n}(L-1) \leq R_{n}(L)$.

	
The opposite direction of Theorem~\ref{th:Ukk} does not always hold. In fact, in~\cite{GabMil18}, an encoding scheme that uses the property from Theorem~\ref{th:Ukk} is used in order to encode $L$-reconstructible strings that are almost $(L-1)$-substring unique. Recently, two encoding schemes of reconstructible binary strings that are also $(L-1)$-substring unique were proposed in~\cite{EliGabMedYaa19}. The first scheme is applied for a window length of $ L = 2\lceil \log (n)\rceil + 2$ with a single bit of redundancy, and the second one works for windows of length $L=\lceil a \log (n) \rceil$ for $ 1 < a \le 2 $ and its asymptotic rate approaches 1. According to the first scheme, one can deduce that $\SubDistSize_{n}(2\lceil \log (n)\rceil + 2) \geq 2^{n-1}$ and the second one implies that for all $ 1 < a \le 2 $, 
$$\lim_{n\rightarrow\infty}\frac{\log_2(\SubDistSize_{n}(\lceil a \log (n) \rceil))}{n} =1.$$ This result is also proved directly in~\cite{EliGabMedYaa19}, by deriving a lower bound on the number of strings in  $\SubDistSet_{n}(\lceil a \log (n) \rceil)$. 

The motivation to study $(L,d)$-substring distant strings originates from the observation that in many cases the $L$-multispectrum cannot be read error-free. This translates to a stronger property, such as the one given by $(L,d)$-substring distant, that strings need to satisfy in order to guarantee unique reconstruction in the presence of errors. 

	
\begin{definition}\label{def:t-losses-ball}
Let $\bfw\in\Sigma^n$ be a string and $S_L(\bfw)$ is its $L$-multispectrum. A multiset $U$ is called a \textbf{$t$-losses $L$-multispectrum of $\bfw$} 
if $U\subseteq S_L(\bfw)$ and $|S_L(\bfw)| - |U|\leq t$. The \textbf{$t$-losses $L$-multispectrum ball of $\bfw$}, denoted by $\cB_{L,t}(\bfw)$, is defined to be the multiset
$$\cB_{L,t}(\bfw) = \{ U \ | \ \textmd{$U$ is a $t$-losses $L$-multispectrum of $\bfw$} \}.$$
\end{definition}

\begin{example}\label{ex2}
	Let $ n, \bfx $ be from Example~\ref{ex1} and let $ L = 8 $.
	The $ L $-multispectrum of $ \bfx $ is 
	\begin{align*}
		S_L(\bfx)& = \{01000001, 10000011, 00000111, 00001110, & \\ 
		& 00011101, 00111011, 01110111, 11101111, 11011111\}. &
	\end{align*}
	The multiset 
	\begin{gather*}
	 U_1 =   \{10000011, 00000111, 00001110, 00011101,\\ 11101111, 11011111\},
	\end{gather*}	 
	which equals to $ S_L(\bfx) \setminus \{ \bfx_{1,L}, \bfx_{5,L}, \bfx_{6,L} \} $, is a 3-losses $L$-multispectrum of $ \bfx $. 
The multiset 
	\begin{gather*}
	U_2 =   \{01000001, 00001110, 00011101, 11101111, \\ 11011111\},
	\end{gather*}	 
	which equals to $ S_L(\bfx) \setminus \{ \bfx_{2,L}, \bfx_{3,L} \bfx_{5,L}, \bfx_{6,L} \} $, is a 4-losses $L$-multispectrum of $ \bfx $. 
\end{example}

Next, we discuss reconstruction from some lossy multispectrum $U$ of a string $\bfw$. Throughout this paper, we assume unless stated otherwise that $ \bfw $ is $ L $-reconstructible and in fact $ L $-substring unique (i.e. all substrings of $ S_L(\bfw) $ are unique) since otherwise the question of reconstructing noisy multispectra is irrelevant.
 
Moreover, notice that if successive substrings of $ \bfw $ are missing from the start or the end of $ U $ then several entries of the input string can be entirely absent from $U$. For example, if the three windows $\bfw_{1,L}, \bfw_{2,L}, \bfw_{n-L+1,L}$ are lost, it will not be possible to restore the values of $w_1,w_2,w_n$ since these entries do not appear at all in $U$. Therefore, given a multispectrum $U$, we define its \textbf{\emph{maximal-reconstructible substring}}, denoted by $\bfW_1(U)$, to be the largest consecutive substring of $ \bfw $ which its $ L $-prefix and $ L $-suffix appear in $ U $ (in our example $\bfW_1(U) = \bfw_{3,n-3}$). Namely, assume $ U = \{\bfw_{i_1,L},\dots,\bfw_{i_m,L} \} $ where $ m \ge n-t $ and $ i_1, \dots, i_m \in [n-L+1] $ are distinct indices, then $ \bfW_1(U) $ is the substring $ (w_{i_1},\dots,w_{i_m+L-1}) $.
It will follow that in our constructions the string $\bfW_1(U)$ is uniquely defined and is independent of the original string $ \bfw $. Since there are at most $t$ losses, it is ensured that the length of $\bfW_1(U)$ is at least $n-t$. Notice that if $ U $ contains the substrings $ \bfw_{1,L}$ and $\bfw_{n-L+1,L} $, it is ensured that $ \bfW_1(U) = \bfw $.

\begin{example}
	Following Example~\ref{ex2}, $ \bfW_1(U_1) = \bfx_{2,15} $ and $ \bfW_1(U_2) = \bfx $.
\end{example}

Accordingly, the following definition presents the family of strings that will be studied in Section~\ref{sec:rec-incomplete}. 
\begin{definition}
A string $\bfw$ is called an \textbf{{$(L,t)$-reconstructible string}}  if for any $t$-losses $L$-multispectrum $U \in \cB_{L,t}(\bfw)$, the maximal-reconstructible substring $\bfW_1(U)$ can be uniquely reconstructed from $ U $.  
\end{definition}

\begin{remark}
Note that in the case of $\bfW_1(U) \neq \bfw $, the reconstruction decoder discovers that it can only reconstruct a part of the string and reconstructs $ \bfW_1(U) $ exactly.
\end{remark}

The string reconstruction problem was motivated by the reading mechanism of DNA sequences. In this process, short substrings are read from the long sequence and are then used to reconstruct the sequence. If the length of the read substrings is $L$, then it is assumed that on each read one of the $n-L+1$ substrings is read with equal probability. The results in the paper can also be used in order to analyze the reduction in the number of required reads while considering the cases of receiving all and almost all of the $n-L+1$ substrings with high probability. 
This will be explained next.

Assume first that the complete $L$-multispectrum $ S_L(\bfx) $ is required to recover the long sequence $\bfx$. Let $C_{\epsilon,0}$ be such that when reading $M=C_{\epsilon,0}n$ substrings of $\bfx$, all $n-L+1$ substrings in  $ S_L(\bfx) $ are read with probability at least $1-\epsilon$. The value of $C_{\epsilon,0}$ can be lower bounded as follows. The probability that a single length-$L$ substring is not read upon $M$ attempts is 
\begin{align*}
	P &= \left(  1- \frac{1}{n-L+1}\right)^M =  \left(  1- \frac{1}{n-L+1}\right)^{C_{\epsilon,0}n} 
	\le 
	e^{-C_{\epsilon,0}}.
\end{align*}
Thus, the probability that upon $M$ reads not all substrings in the $L$-multispectrum $ S_L(\bfx) $ are read can be upper bounded by the union bound as 
$$P_0 \leq (n-L+1) P =  (n-L+1)  e^{- C_{\epsilon,0}} \le ne^{- C_{\epsilon,0}}.$$
Hence, in order to guarantee success probability of at least $1-\epsilon$, it suffices that $ne^{- C_{\epsilon,0}} \leq \epsilon$, i.e.,
$$ C_{\epsilon,0} \geq \ln( n) +\ln(1/\epsilon).$$

On the other hand, if it is possible to reconstruct the sequence $\bfx$ even in the presence of $t$ losses, then a reconstruction failure occurs when at least $t+1$ substrings are not read. Hence this probability is given by 
$$P_t \leq  \binom{n-L+1}{t+1} \cdot P^{t+1} \leq (ne^{- C_{\epsilon,t}})^{t+1},$$
where now the required number of reads is $C_{\epsilon,t}n$. Accordingly, in order to guarantee $1-\epsilon$ success probability, it is enough to require that $ (ne^{- C_{\epsilon,t}})^{t+1} \leq \epsilon$ and hence
$$C_{\epsilon,t} \geq \ln( n) +\frac{\ln(1/\epsilon)}{t+1}.$$
Hence, if for example $1/\epsilon = \cO(n^a)$, then the number of reads can be reduced roughly by a factor of $(t+1)(1-\frac{t}{a+t+1})$.

Lastly, for the convenience of the reader, relevant notations and terminology referred to throughout the paper are summarized in Table~\ref{tab:cql}.
\begin{table*}[!h]
	\caption{Table of Definitions and Notations}
	\begin{center}		
		\label{tab:cql}	\begin{tabular}{cccc}\hline
			Notation & Meaning & Remarks   \\
			\hline
$ n $ & String length & Sec.~\ref{sec:def}  \\
$ \Sigma $ & The binary alphabet $ \{0,1\} $ & Sec.~\ref{sec:def}  \\
$ \bfw $ & A length-$ n $ string & Sec.~\ref{sec:def}  \\
$ \cB_{r}(\bfw) $ & The radius-$ r $ ball around $ \bfw $ & Sec.~\ref{sec:def}  \\
$ L $ & Substring length & Sec.~\ref{sec:def}  \\
$ S_L(\bfw) $ & $ L $-multispectrum of $ \bfw $ & Sec.~\ref{sec:def}  \\
$ \SubDistSet_{n}(L,d)$ & The set of $ (L,d) $-substring distant length-$ n $ strings & Sec.~\ref{sec:def}  \\
$ \SubDistSize_{n}(L,d)$ & The size of $ \SubDistSet_{n}(L,d)$ & Sec.~\ref{sec:def}  \\
$ \cR_n(L)$ & The set of $ L $-reconstructible length-$ n $ strings & Sec.~\ref{sec:def}  \\
$ R_{n}(L,d)$ & The size of $\cR_n(L)$ & Sec.~\ref{sec:def}  \\
$ \cB_{L,t}(\bfw)$ & The $t$-losses $L$-multispectrum ball of $\bfw$ & Def.~\ref{def:t-losses-ball}  \\
$ \bfW_1(U)$ & The maximal-reconstructible substring of a lossy multispectrum $ U $ & Sec.~\ref{sec:def}  \\
$ \ell_1 $ & The value $L - \floorenv{t/3} -1 $ & Sec.~\ref{sec:rec-incomplete}  \\
$ \ell_2 $ & The value $L - \ceilenv{2t/3} -1 $ & Sec.~\ref{sec:rec-incomplete}  \\
$ \ell_3 $ & The value $L-t-1 $ & Sec.~\ref{sec:rec-incomplete}  \\
$ I_2 $ & The set $ [n- \ell_2 - t+1,n- \ell_2 + 1] $ & Sec.~\ref{sec:rec-incomplete}  \\
$ I_3 $ & The set $ [n- \ell_3 - t+1,n- \ell_3 + 1] $ & Sec.~\ref{sec:rec-incomplete}  \\
$ \cD_n(L,t) $ & The set of length-$ n $ strings that satisfy $(n,L,t)$-LREC constraints & Sec.~\ref{sec:rec-incomplete}  \\
$ D_n(L,t) $ & The size of $ \cD_n(L,t) $ & Sec.~\ref{sec:rec-incomplete}  \\
$ \cB_{L,t,s}(\bfw) $ & The $(t,s)$-erroneous $L$-multispectrum ball of $\bfw$ & Def.~\ref{def:ts-erroneous}  \\
$ \bfW_2(U)$ & The majority maximal-reconstructible substring of an erroneous multispectrum $ U $ & Sec.~\ref{sec:rec-Erroneous}  \\
$ \bfu_d $ & A $ d $-auto cyclic string & Sec.~\ref{sec:rec-Erroneous}  \\
$ \cC\cB_{n,r}(\bfw) $ & The concatenation ball of radius-$ r $ around $ \bfw $ & Sec.~\ref{sec:rec-Erroneous}  \\
$ \cD_n(L,t,s) $ & The set of length-$ n $ strings that satisfy $(n,L,t,s)$-EREC constraints & Sec.~\ref{sec:rec-Erroneous2}  \\
$ D_n(L,t,s) $ & The size of $ \cD_n(L,t,s)$ & Sec.~\ref{sec:rec-Erroneous2}  \\
$ \bfW_3(U)$ & The consensus maximal-reconstructible substring of an erroneous multispectrum $ U $ & Sec.~\ref{sec:rec-Erroneous2}  \\
$ I(V) $ & The set $[i_1,i_m + L-1]$ for $ V = \{\bfu_{i_1},\dots, \bfu_{i_m}\}$  & Sec.~\ref{sec:rec-Erroneous2}  \\
$ n(V) $ & The size of $ I(V) $  & Sec.~\ref{sec:rec-Erroneous2}  \\
$ SB_{k}(\bfw)  $ & The $ k $-striping ball of $ \bfw $ & Sec.~\ref{sec:rec-Erroneous2}  \\
			\hline
		\end{tabular}
	\end{center}
\end{table*}

	\section{Reconstructing an Incomplete Multispectrum}\label{sec:rec-incomplete}
	In this section, we define constraints for $(L,t)$-reconstructible strings, propose a reconstruction algorithm for those strings and analyze the cardinality of such family of strings.  
\subsection{Reconstruction Constraints}

The goal of this subsection is to construct $t$-losses $L$-reconstructible strings. This will be given by strings that satisfy a few constraints, given in the next definition. For simplicity, we consider here only the binary case, so 
$\Sigma=\{0,1\}$. 

For the rest of this section, we denote the integers $ \ell_1 = L - \floorenv{t/3} -1, \ell_2 =  L - \ceilenv{2t/3} -1, \ell_3 = L-t-1 $ and the sets $ I_2 = [n- \ell_2 - t+1,n- \ell_2 + 1], I_3 = [n- \ell_3 - t+1,n- \ell_3 + 1]$.

\begin{definition}\label{def:reconstruction-constratints} 
A string $\bfx\in\Sigma^n$ is said to satisfy the \textbf{$(n,L,t)$-lossy reconstruction (LREC) constraints} if it fulfills the following three constrains. 
\begin{enumerate}
\item $ \bfx$ is a $ \ell_1$-substring unique string.
\item The first and last $ t +1 $ length-$ \ell_2$ substrings are not identical to all other length-$ \ell_2$ substrings. Namely, for all $ i \in [t \plus 1], j \in [n- \ell_2 + 1]$ with $ i \neq j $ then	$ \bfx_{i,\ell_2} \neq  \bfx_{j,\ell_2}$ and for all $i \in [n-\ell_2+1], j \in I_2$ with $ i \neq j $, then $ \bfx_{i,\ell_2} \neq  \bfx_{j,\ell_2}$.
\item The first $ t + 1 $ length-$\ell_3$ substrings are not identical to the last $ t + 1 $ length-$\ell_3$ substrings. Namely, for all $ i \in [t \plus 1], j \in I_3 $, $\bfx_{i,\ell_3} \neq  \bfx_{j,\ell_3}$.
\end{enumerate}
\end{definition}


For $n,L,t$, denote by $\cD_n(L,t)$ the set of all strings that satisfy the $(n,L,t)$-LREC constraints and let $D_n(L,t) = |\cD_n(L,t)|$. Note that by definition, if a string satisfies the $(n,L,t)$-LREC constraints it satisfies the $(n,L,t')$-LREC constraint for all $t'\leq t$, that is, $\cD_n(L,t) \subseteq  \cD_n(L,t')$.
\begin{example}
	Let $ n, L, \bfx $ from Example~\ref{ex2}. The string $ \bfx $ satisfies the $ (n,L,4) $-LREC constraints. The first constraint follows from the fact that $ \bfx $ is $ 6 $-substring unique and it is possible to verify that the two other constraints are satisfied as well. Therefore, $ \bfx \in \cD_n(L,4) $ and also  $ \bfx \in \cD_n(L,3) $.
\end{example}

In \cite{GabMil18}, the authors focused on a type of errors which corresponds to occurrence of bursts of substring losses. They identified a lossy multispectrum $ U \subseteq S_L(x) $ to have $ G $-maximal coverage gap if $ G $ is the maximum number of consecutive substrings that are not included in $ S_L(\bfx) $. Based on this characterization, they showed that if $ \bfx $ is $ (L-G-1) $-substring unique it is reconstructible from such a lossy multispectrum $U$. When applying this constraint to our problem, assume that $ U \in \cB_{L,t}(\bfx) $, then it is necessary that $ G = t $ since all the losses can occur consecutively. Hence, the construction proposed in \cite{GabMil18} requires that $ \bfx $ is $ (L-t-1) $-substring unique in order for it to be  $ (L,t) $-reconstructible. 
Based on the results of \cite{EliGabMedYaa19}, in order to construct a rate-$(1 - o(1))$ code of $ (L-t-1) $-substring unique length-$ n $ strings, it is required that $ L > a\log (n)+t $ for some $a>1$. It will be shown in Section~\ref{chp:cardinality} that the $ (n,L,t) $-LREC constraint composes a rate-$(1 - o(1))$ code for values of $L$ that satisfies $ L > a\log (n)+t/3 $, where $a>\max\{b/3+1,2b/3\}$ and $t=b\log (n) + o(\log (n))$ for $ b \in \mathbb{N} $. Hence, for these parameters, the construction proposed in this paper imposes a weaker constraint on the value of $L$ than the construction proposed in \cite{GabMil18}.

Moreover, in~\cite{ChaChrEzeKia17}, the authors studied the problem of reconstructing an incomplete multispectrum under the name of \emph{assembly of words in presence of coverage errors}. They presented constructions with an asymptotic rate that is strictly less than $ 1 $, that achieve the following results: 1. For $ L = c \log(n) $, where $ c > 1 $, the construction is able to tolerate $(c-2) \log(n)$ losses but the asymptotic rate approaches $(1-2/c)$, and 2. For $L = cn$ with $0<c<1$, the construction is able to tolerate $cn - o(n) $ losses with an asymptotic rate that approaches $1-2c$. Thus, for $t=b\log (n) + o(\log (n))$ losses where $ b \in \mathbb{N} $, the construction proposed in this paper has higher rate than the one proposed in~\cite{ChaChrEzeKia17} and requires smaller $ L $.

\subsection{Reconstruction Algorithm}

Our next goal is showing that every string which satisfies the $(n,L,t)$-LREC constraint is an $(L,t)$-reconstructible string, that is, its maximal-reconstructible substring can be uniquely decoded even if at most some $t$ substrings are not read. Namely, we prove the following theorem.
\begin{theorem}\label{th:LREC reconstruction}
Every string $\bfx \in \cD_n(L,t)$ is an $(L,t)$- reconstructible string.
\end{theorem}

The proof of Theorem~\ref{th:LREC reconstruction} is given by an explicit decoding algorithm which receives a multiset $U\in \cB_{L,t}(\bfx)$ for some $\bfx\in \cD_n(L,t)$. First, we present in Algorithm~\ref{alg:Stitch}, an auxiliary procedure, called the \emph{Stitching Algorithm}, which receives two inputs: 1) A set $ A $ of substrings that we aim to stitch, and 2) $ \rho \le t $, a parameter that will indicate the minimum overlapping size of two substrings in order to be stitched together. The stitching algorithm is based on iterative stitching steps and is composed of three nested loops. At the most inner loop, two substrings are stitched if the suffix of the first is identical to the prefix of the second. This will later indicate that these substrings originated from the same positions in the input string. The middle loop constructs continuous substrings of  $ U $ by finding a prefix of such a substring and repeatedly applying the inner loop in order to correctly concatenate to it more bits. The outer loop iterates over $ k = 0, \dots, \rho $ and at every iteration we bridge gaps that were created by losses of $ k $ consecutive substrings. This is accomplished by reducing the substring length used at the suffix-prefix matching condition method of the inner loop. The stitching algorithm returns  a set of continuous substrings reconstructed from $ U $. which its size is smaller than the input set size, or equal if no stitching occurred. We say that an operation of the stitching algorithm is \emph{successful} if the output set size is strictly smaller than the input set size. 

\begin{algorithm}
	\caption{Stitch($ A,\rho $)}\label{alg:Stitch}
	\begin{algorithmic}[1]				
		\For{$ k = 0, \dots, \rho $ }
		\State{$ B = \emptyset $}
		\While{$ A \neq \emptyset $}
		\State{pick $ \bfw \in A $ such that for every other $ \bfw' \in A $, $ $Pref$_{L-k-1}(\bfw) \neq $Suff$_{L-k-1}(\bfw') $ }\label{step:pick-unique}
		\State{set $ A = A\setminus \{\bfw\}$}
		\While{there exists $ \bfw' \in A $ such that Suff$_{L-k-1}(\bfw) = $ Pref$_{L-k-1}(\bfw') $} \label{step:match}
		\State{set $ \bfw = \bfw \circ $ Suff$_{|\bfw'|-L+k+1}(\bfw') $}\label{step:concat}
		\State{set $ A = A\setminus \{\bfw'\}$}\label{step:A_update}
		\EndWhile
		\State{$ B = B \cup \{\bfw\} $}
		\EndWhile
		\State{$ A = B $}
		\EndFor
		\State{\textbf{return} $ B $}		
	\end{algorithmic}
\end{algorithm}

Algorithm~\ref{alg:reconstruct-losses}, called the \emph{Reconstruction Algorithm}, receives a $t$-losses $ L- $multispectrum $ U $ for some $ \bfx \in \cD_n(L,t)$ and uses the stitching algorithm to reconstruct  $\bfW_1(U)$, the maximal reconstructible substring of $U $. In case the returned set by the reconstruction algorithm consists of a single string we assume that the output is the string itself (i.e. not a set).
\begin{algorithm}
	\caption{Reconstruct($U,t$)}\label{alg:reconstruct-losses}
	\begin{algorithmic}[1]
		\Require  $ U\in \cB_{L,t}(\bfx) $ for some $ \bfx \in \cD_n(L,t)$
		\Ensure $ \bfW_1(U) $ the maximum reconstructible-substring of $ U$
		\State{Invoke $ A_0 =$  Stitch$(U,\floorenv{t/3}) $. } \label{step:rec-loss}
		\State{If $ |A_0| = 1$ and $A_0=\{\bfy\}$: return $\bfy$.} \label{step:rec-loss1}
		\State{If $|A_0|=2$ and $A_0=\{\bfy_1,\bfy_2\}$: return Stitch$(A_0,t) $. }\label{step:rec-loss-2}
		\State{If $ |A_0| = 3 $ and $A_0=\{\bfy_1,\bfy_2,\bfy_3\}$: for $i=1,2,3$ invoke $ A_i =$ Stitch$(A_0 \setminus \{\bfy_i\},\ceilenv{2t/3})$ and if successful invoke $A_i'=$ Stitch$(A_i \cup \{\bfy_i\},\ceilenv{2t/3}) $. If successful again, return $A_i'$.}\label{step:rec-loss-3}
	\end{algorithmic}	
\end{algorithm}

Assume that $U= \{\bfx_{i_1,L},\ldots,\bfx_{i_m,L}\}$, where $1\leq i_1<i_2<\cdots < i_m\leq n-L+1$. Let $ A_0 =$  Stitch$(U,\floorenv{t/3}) $ be the resulting set after Step~\ref{step:rec-loss}, and denote $A_0= \{\bfy_1,\ldots,\bfy_r\}$. Additionally, denote by $ B_k $ the set $ B $ after the $ k $-th iteration of the for loop of Algorithm~\ref{alg:Stitch}. Note that every substring in $U$ is a substring of \emph{exactly} one string from $A_0$, and that $ A_0 = B_{\floorenv{t/3}} $. The next two examples demonstrate how Algorithms~\ref{alg:Stitch} and~\ref{alg:reconstruct-losses} operate. 

\begin{example}
	Let $ n, L, \bfx,U_1 $ from Example~\ref{ex2}, so that $ U_1 \in \cB_{L,t}(\bfx)$, where $t=3$. Assume that we invoke Reconstruct$(U_1,t)$. First, the algorithm invokes $A_0=$ Stitch($ U_1, 1$). At the first iteration of the for loop where $k=0$, assume the algorithm picks $ \bfx_{2,8} = 10000011 $ and stitches to it $ \bfx_{3,8} = 00000111 $ followed by $ \bfx_{4,8} = 00001110 $. Next, the algorithm picks $ \bfx_{7,8} = 01110111 $ and stitches to it $ \bfx_{8,8} = 11101111$ followed by $ \bfx_{9,8} = 11011111 $. Thus, we have at the end of this iteration
	\[
		B_0 = \{\bfx_{2,10}, \bfx_{7,10}\} = \{ 1000001110, 0111011111 \}. 
	\]
	No stitching is made at the second iteration for $k=1$ and thus $ A_0 = B_1 = B_0 $ is the output of the stitching algorithm. Since $ |A_0|=2 $, we execute next in Step~\ref{step:rec-loss-2}, Stitch($ A_0, 3 $). Then, the two substrings of $A_0$ are stitched at iteration $ k=2 $, since Suff$ _5(\bfx_{2,10}) =  $ Pref$ _5(\bfx_{7,10}) $. Eventually, the string 
	\[
		\bfx_{2,15} = 1000001110111 = \bfW_1(U_1)
	\]
	is returned as expected.
\end{example}

\begin{example}
	Let $ n, L, \bfx, U_2 $ from Example~\ref{ex2}, so that $ U_2 \in \cB_{L,t}(\bfx)$, where $t=4$. Invoking Stitch($ U_2, 1$) returns 
	\begin{align*}
	A_0 = B_1=B_0 =&\{\bfx_{1,8}, \bfx_{4,8}, \bfx_{7,10} \} \\=&\{ 10000011,00011101,0111011111 \}.
	\end{align*}	
	Since $ |A_0|=3 $, the reconstruction algorithm executes Step~\ref{step:rec-loss-3}. Assume that $\bfy_1 = \bfx_{4,8}, \bfy_2 = \bfx_{1,8}$, and $\bfy_3 = \bfx_{7,10}$. For $i=1$, the algorithm receives that Stitch($ A_0 \setminus \{\bfy_1\}, 3$) $=A_0 \setminus \{\bfy_1\}$ which yields with an unsuccessful result. However, for $i=2$, when invoking Stitch($ A_0 \setminus \{\bfy_2\}, 3$), the algorithm stitches the substrings $\bfy_1,\bfy_3$ at iteration $ k=2 $ using Suff$ _5(\bfx_{1,8}) =  $ Pref$ _5(\bfx_{4,8}) $, and returns 
	\[ A_2 = \{\bfx_{1,11}\} = \{ 10000011101 \}. \]
	Lastly, the algorithms applies Stitch($ A_2 \cup \{\bfy_2\}, 3$) and stitches again at iteration $ k=2 $ to receive $ A_2' = \bfx = \bfW_1(U_2) $ as the final result.
\end{example}

The correctness of Algorithms~\ref{alg:Stitch} and~\ref{alg:reconstruct-losses} is proved in the next few claims. 

\begin{claim}\label{claim:same-substring}
For all $1\leq j\leq m-1$ if $i_{j+1} - i_j \leq \floorenv{t/3} + 1$, then $\bfx_{i_j,L}$ and $\bfx_{i_{j+1},L}$ are substrings of the same string in $A_0$.
\end{claim}
\begin{IEEEproof}
First, from Constraint 1 it follows that
\begin{align}\label{eq:cons1}
\forall k \in [0,\floorenv{t/3}], \bfx  \text{ is }  (L-k-1) \text{-substring unique.}
\end{align}
We first claim that for every $ k < k' =  i_{j+1} - i_j - 1 $, $\bfx_{i_j,L},\bfx_{i_{j+1},L}$ are not substrings of the same string in $B_k$ and furthermore, $\bfx_{i_j,L}$ is a suffix of some substring $ \bfw_{1} \in B_k $, while $\bfx_{i_{j+1},L}$ is a prefix of another substring $ \bfw_{2} \in B_k $. To see this, assume in the contrary and let $ k_0 < k' $ be the first iteration where $\bfx_{i_j,L}$ is not a suffix of some substring in $ B_{k_0} $. That is, there exists $ \bfw \in B_{k_0-1}$ where $ \text{Suff}_{L}(\bfw)=\bfx_{i_j,L} $, that is stitched to the left of another $\bfw' \in B_{k_0-1}$, which satisfies $ \text{Pref}_{L}(\bfw')=\bfx_{i_g,L} $ for another substring $ \bfx_{i_g,L} \in U $. However, it follows from (\ref{eq:cons1}) that $ i_j < i_g < i_{j+1} $ and therefore such a substring cannot exist in $ U $. 

In particular, $\bfx_{i_j,L},\bfx_{i_{j+1},L}$ are not substrings of the same string in $B_k$. Thus, at the $k'$-th iteration, the substrings are stitched since
\begin{align*}
\text{Suff}_{L-k'-1}(\bfw_{1}) &=  \bfx_{i_j + k'+1,L-k'-1} =\bfx_{i_{j+1},L-k'-1}  \\&= \text{Pref}_{L-k'-1}(\bfw_{2}).
\end{align*}
Note that from (\ref{eq:cons1}), for any other substring $ \bfw_{3} \not \in \{\bfw_{1},\bfw_{2}\}$, $   \text{Suff}_{L-k'-1}(\bfw_{1})$ $\neq \text{Pref}_{L-k'-1}(\bfw_{3}) $ and $   \text{Pref}_{L-k'-1}(\bfw_{2}) \neq \text{Suff}_{L-k'-1}(\bfw_{3}) $.

\end{IEEEproof}	

It is said that a spectrum $U$ experienced a \emph{burst of losses} of length $ h $ at index $ i \le n-L-t+1 $ if $ \bfx_{i,L}, \ldots, \bfx_{i+h-1,L} \not \in U $. 
\begin{claim}\label{claim:first-op}
The set $A_0 = $ Stitch$(U,\floorenv{t/3}) $ satisfies $ |A_0| \le 3 $. 
\end{claim}	
\begin{IEEEproof}
Following Claim~\ref{claim:same-substring}, we will show that there are three possible cases for the size of the set $A_0$. First, if there are no bursts of losses longer than $\floorenv{t/3}$, then all substrings of $ U $ are contained in a single string of $ A_0 $, thus $ |A_0| = 1 $. Second, if there is a single burst of losses longer than $\floorenv{t/3}$, then the substrings of $ U $ are divided into two different strings of $ A_0$, thus $ |A_0| = 2 $. Similarly, at the third case there are two bursts of losses longer than $\floorenv{t/3}$, and then $ |A_0| = 3 $. Other cases are not possible, since the number of losses is at most $ t $.
\end{IEEEproof}	

\begin{claim}\label{claim:second-op}
At Step~\ref{step:rec-loss-2} of Algorithm~\ref{alg:reconstruct-losses}, the result of Stitch$(A_0,t)$ is $\bfW_1(U)$.
\end{claim}	
\begin{IEEEproof}
Let $ A_0 = \{\bfy_1,\bfy_2\} $, where Suff$_L(\bfy_1) = \bfx_{i_j,L} $ and Pref$_L(\bfy_2) = \bfx_{i_{j+1},L} $. It follows that Pref$_L(\bfy_1) = \bfx_{i_1,L} $, where $i_1\leq t+1$. Therefore, the prefix Pref$_{L-t-1}(\bfy_1)$ is one of the first $ t+1 $ length-$(L-t-1)$ substrings of $ \bfx $. Similarly, Suff$_L(\bfy_2) = \bfx_{i_m,L} $, where $i_m\geq n-L-t+1$, and the suffix Suff$_{L-t-1}(\bfy_2)$ is one of the last $ t+1 $ length-$(L-t-1)$ substrings of $ \bfx $. Thus, from Constraint 3, for every $ k \in [0,t] $, Pref$_{L-k-1}(\bfy_1) \neq $ Suff$_{L-k-1}(\bfy_2)$. Therefore, it is not possible to stitch the substring $\bfy_1$ to the right of $ \bfy_2$. Since there are at most $ t $ losses, it follows that $ i_{j+1} - i_j - 1 \le t $. Hence, these substrings are stitched correctly to a single string at iteration $k' =  i_{j+1} - i_j - 1$, which results with the string $\bfW_1(U)$.
\end{IEEEproof}

\begin{claim}\label{claim:third-op}
At Step~\ref{step:rec-loss-3} of Algorithm~\ref{alg:reconstruct-losses}, there exists a substring  $ \bfy_i \in A_0 $ such that both operations of the stitching algorithm are successful. For such a $ \bfy_i$,  the result of this step is the string $\bfW_1(U)$.
\end{claim}	
\begin{IEEEproof}
Let $ A_0 = \{\bfy_1, \bfy_2, \bfy_3\} $ where Suff$_L(\bfy_1) = \bfx_{i_j,L}$,  Pref$_L(\bfy_2) = \bfx_{i_{j+1},L} $,  Suff$_L(\bfy_2) = \bfx_{i_h,L}$,  and Pref$_L(\bfy_3) = \bfx_{i_{h+1},L} $. Since the number of losses is at most $t$, it follows that  $ \floorenv{t/3} < i_{j+1} - i_j -1 \le \ceilenv{2t/3}  $ and $ \floorenv{t/3} < i_{h+1} - i_h - 1 \le \ceilenv{2t/3}  $.
Similarly to the proof of Claim~\ref{claim:second-op}, but in this case according to Constraint 2, for every $ k \in [0,\ceilenv{2t/3}] $,
\begin{align}\label{eq:cons2_1}
\text{Pref}_{L-k-1}(\bfy_1) \neq  \text{ Suff}_{L-k-1}(\bfy_s) \text{ for } s \in [2,3],
\end{align}
\begin{align}\label{eq:cons2_2}
\text{Suff}_{L-k-1}(\bfy_3) \neq  \text{ Pref}_{L-k-1}(\bfy_s) \text{ for } s \in [1,2].
\end{align} 
Thus, if we pick $ \bfy_1 $, from (\ref{eq:cons2_2}) we can only stitch $ \bfy_2 $ to the left of $ \bfy_3 $ at Stitch$(A_0 \setminus \{\bfy_1\},\ceilenv{2t/3}) $ and from (\ref{eq:cons2_1}) we stitch the result to the right of $ \bfy_1 $ at Stitch$(A_1 \cup \{\bfy_1\},\ceilenv{2t/3}) $. The result is similar if we initially pick $ \bfy_3 $.
If we pick $\bfy_2 $ it follows that it is only possible to incorrectly stitch $ \bfy_1$ to the left of $\bfy_3 $ at Stitch$(A_0 \setminus \{\bfy\},\ceilenv{2t/3}) $. However, it is ensured from (\ref{eq:cons2_1}) and (\ref{eq:cons2_2}) that the resulting string in this case cannot be stitched to $ \bfy_2 $ at the second operation of the stitching algorithm.
	
Since both operations are successful, the result contains a single substring which contains all the substrings of $ U $.
	
\end{IEEEproof}

\begin{lemma}\label{lem:alg-reconstruct-incomplete}
	Algorithm~\ref{alg:reconstruct-losses} uniquely reconstructs $\bfW_1(U)$.  
\end{lemma}
\begin{IEEEproof}
	Following Claim~\ref{claim:first-op}, there are three possible cases for the size of $A_0$. From Claims~\ref{claim:same-substring}, \ref{claim:second-op}, and \ref{claim:third-op}, Algorithm~\ref{alg:reconstruct-losses} returns a single string where all the elements of $ U $ are a substrings of it. That is the maximum reconstructible-substring of $ U $. 
\end{IEEEproof}

Lemma~\ref{lem:alg-reconstruct-incomplete} verifies also the proof of Theorem~\ref{th:LREC reconstruction}.

\subsection{Cardinality Analysis}\label{chp:cardinality}

Our next goal is to estimate the value of $D_n(L,t)$ for some specific parameters of $n,L,t$. 
Our approach is based on the probabilistic method used in~\cite{EliGabMedYaa19IEEE} to prove that the asymptotic rate of the set $\SubDistSet_{n,2}(L)$ approaches 1, when $L = \ceilenv{a \log (n)}$ and $a>1$. Building upon this approach, for a given value of $t$ that satisfies $t=b\log(n) +o(\log(n))$ for some $ b \in \mathbb{N} $, we show how to choose the value of $L$ such that the three $ (n,L,t) $-LREC constraints hold. This result is proved in the following theorem.
\begin{theorem}\label{th:D}
Let $t=b\log(n) +o(\log(n))$ for $ b \in \mathbb{N} $ and $ L = \ceilenv{a \log (n)} + \floorenv{t/3} + 1 $ where $a > \max \{b/3+1, 2b/3 \}$, then it holds that 
$$\lim_{n\rightarrow\infty}\frac{\log(D_{n}(L,t))}{n} =1.$$
\end{theorem}
We follow a similar outline as the proof presented in~\cite{EliGabMedYaa19IEEE} for $ L $-substring unique strings, and prove Theorem~\ref{th:D} using the asymmetric Lo\`asz local lemma which was first proved in~\cite{ErLo73} and is stated next as it appears in~\cite{AlSp2000}. 
\begin{lemma}[\cite{AlSp2000}, Lemma~5.1.1]~\label{lem:lll}
	Let $ Y_0, \dots, Y_{m-1} $ be events in the arbitrary probability space. Let $ G = (V,E) $ be a graph with $ V = [m] $ such that for every $ i \in [m] $, the event $ Y_i $ is mutually independent of all the events $ \{Y_j \mid (i,j) \not \in E \} $. Suppose  that there are real numbers $ \alpha_0, \dots, \alpha_{m-1} $ such that $ \alpha_i \in [0,1] $ and for all $ i \in [m] $, $$ Pr(Y_i) \le \alpha_i \prod_{(i,j)\in E}(1-\alpha_j). $$ 
	Then, it is satisfied that $$ Pr \left(  \bigcap_{i \in [m]} \overline{Y}_{i}\right) \ge \prod_{i \in [m]}(1-\alpha_i) $$
	where $\overline{Y}_{i}$ is the complement of $ Y_i $.
\end{lemma}

\begin{IEEEproof}[Proof of Theorem~\ref{th:D}]
For the values of $t$ and $L$ stated in the theorem it holds that 
$ L = (a +b/3) \log (n) + o(\log(n))$,  $\ell_1 = a \log (n) + o(\log(n))$,  $\ell_2 = (a-b/3) \log (n) + o(\log(n))$, and $\ell_3 = (a-2b/3) \log (n) + o(\log(n))$. Note that for simplicity, we assume all are integer values.
The size of $ \cD_n(L,t) $ will be estimated by a probabilistic approach. Assume that $ \bfw $ is a length-$ n $ string chosen uniformly at random over $ \Sigma $. In order to estimate the value of $ D_n(L,t) $, we may estimate the probability $ Pr(\bfw \in \cD_n(L,t)) $ since $ D_n(L,t) = 2^n \cdot Pr(\bfw~\in~\cD_n(L,t))  $ and hence
\begin{equation}\label{eq:da-rate}
	\begin{aligned}
		\lim_{n\rightarrow\infty}\frac{\log(D_{n}(L,t))}{n}
		= 1 + \limup{n}\frac{1}{n}\log(Pr(\bfw \in \cD_n(L,t))).
	\end{aligned}
\end{equation}

For $ \ell \in \{ \ell_1,\ell_2,\ell_3 \} $ and positions $ i, j \in [n-\ell+1] $ we notate $ \bfz = (i,j,\ell) $ and denote by $ I_{\bfz} = \mathbbm{1}(\bfw_{i,\ell} =\bfw_{j,\ell})$ the indicator function of the event that the $ \ell $-substrings that start at positions $ i $ and $ j $ are identical. According to the LREC constraints, we denote
\begin{align*}
	&\Gamma_1 = \{ (i,j,\ell_1) \mid  i,j \in [n-\ell_1 +1] \text{ and } i<j \},
	\\&\Gamma_2 = \left\{ (i,j,\ell_2) \middle\vert  \begin{array}{l} i \in [t+1], j \in [n-\ell_2+1], i<j \text{ or } \\ i \in [n-\ell_2 +1], j \in I_2, i<j \end{array}\right\},	\\
	&\Gamma_3 = \{ (i,j,\ell_3) \mid  i \in [t+1], j \in I_3 \},
\end{align*}
and let $ \Gamma = \Gamma_1 \cup \Gamma_2 \cup \Gamma_3 $ denote the set of all admissible triples. Note that $ |\Gamma_1| \le n^2, |\Gamma_2| \le 2(t+1)n, $ and $ |\Gamma_3| \le (t+1)^2 $. We are interested in a lower bound on 
$$ Pr(\bfw \in \cD_n(L,t)) = Pr \left(\sum_{\bfz \in \Gamma} I_{\bfz} = 0\right). $$

Note that for every $ \bfz = (i, j, \ell) \in \Gamma $ it holds that $Pr(I_\bfz) = \frac{1}{2^\ell} $. Let $ \bfz = (i, j, \ell), \bfz' = (i', j', \ell') \in \Gamma  $. It is clear that if the substrings $ \bfw_{i,\ell}, \bfw_{j,\ell} $ do not overlap with $ \bfw_{i',\ell'}$ or $ \bfw_{j',\ell'} $, then the indicators $ I_{\bfz}, I_{\bfz'} $ are independent. We use Lemma~\ref{lem:lll} with a graph $ G = (V,E) $ such that $ V =  \Gamma $ and there is an edge $ \bfz \rightarrow \bfz' $  if at least one of $ \bfw_{i,\ell}, \bfw_{j,\ell} $ overlaps with $ \bfw_{i',\ell'}$ or $ \bfw_{j',\ell'} $. It follows that every $ \bfz = (i,j,\ell)  \in \Gamma $ has at most $ 2(\ell_1+\ell)n $ neighbors from $ \Gamma_1 $ in the graph.

We set the numbers $ \alpha_\bfz = \frac{1}{4\ell_1n} $ for every $ \bfz \in \Gamma_1 $, $ \alpha_\bfz = \frac{1}{2(t+1)n} $ for every $ \bfz \in \Gamma_2 $ and $ \alpha_\bfz =  \frac{1}{(t+1)^2} $ for $ \bfz \in \Gamma_3 $. For every $ \bfz \in \Gamma $  it holds that
\begin{equation}\label{eq:a1}
\begin{aligned}
&\prod_{(\bfz, \bfz') \in E} (1-\alpha_{\bfz'}) 
\\&\ge \prod_{(\bfz, \bfz') \in E : \bfz' \in \Gamma_1} (1-\alpha_{\bfz'}) \prod_{\bfz' \in \Gamma_2} (1-\alpha_{\bfz'}) 
\prod_{\bfz' \in \Gamma_3} (1-\alpha_{\bfz'}) 
\\&\geq \left(1\hspace*{-0.5ex}-\hspace*{-0.5ex}\frac{1}{4\ell_1n } \right)^{4\ell_1n }\hspace*{-0.5ex} \left(1\hspace*{-0.5ex}-\hspace*{-0.5ex}\frac{1}{2(t\hspace*{-0.5ex}+\hspace*{-0.5ex}1)n } \right)^{2(t+1)n }\hspace*{-0.5ex} \left(1\hspace*{-0.5ex}-\hspace*{-0.5ex}\frac{1}{(t\hspace*{-0.5ex}+\hspace*{-0.5ex}1)^2} \right)^{(t+1)^2 } 
\\&\ge \frac{1}{e^3},
\end{aligned}
\end{equation}
since each of the expressions in the last inequality approaches $ e^{-1} $ from above as {$ n~ \rightarrow~\infty $}. The condition of the lemma holds for every $ \bfz \in \Gamma $ since
\begin{align*}
\forall_{\bfz \in \Gamma_1} :  Pr(I_\bfz) &= \frac{1}{2^{\ell_1}} = \frac{1}{n^{a}} \overset{\text{(i)}}{\le} \frac{1}{4\ell_1n} \cdot \frac{1}{e^3} \\&\le \alpha_\bfz \prod_{(\bfz, \bfz') \in E} (1-\alpha_{\bfz'}),
\end{align*}
\begin{align*}	
\forall_{\bfz \in \Gamma_2} :  Pr(I_\bfz) &= \frac{1}{2^{\ell_2}} = \frac{1}{n^{a-b/3}}  \overset{\text{(ii)}}{\le}  \frac{1}{2(t+1)n} \cdot \frac{1}{e^3} 
\\&\le \alpha_\bfz \prod_{(\bfz, \bfz') \in E} (1-\alpha_{\bfz'}),
\end{align*}
\begin{align*}	
\forall_{\bfz \in \Gamma_3} :  Pr(I_\bfz) &= \frac{1}{2^{\ell_3}} = \frac{1}{n^{a-2b/3}}  \overset{\text{(iii)}}{\le}  \frac{1}{(t+1)^2} \cdot \frac{1}{e^3} \\&\le \alpha_\bfz \prod_{(\bfz, \bfz') \in E} (1-\alpha_{\bfz'}),
\end{align*}
where  (i) follows from $ 4\ell_1 n = o(n^a) $ since $ a > 1 $, (i)  follows from $ 2(t+1)n = o(n^{a-b/3}) $ since $ a - b/3 > 1  $, $ t = b \log (n) + o(\log(n)) $ and  (iii)  follows from $ (t+1)^2 = o(n^{a-2b/3}) $ since $ t = b \log (n) + o(\log(n)) $ and $ a -2b/3 > 0 $.

By applying Lemma~\ref{lem:lll} and from equation (\ref{eq:a1}) we obtain
\begin{align*}
	Pr(\bfw \in \cD_n(L,t)) &\ge \prod_{\bfz \in \Gamma}\left(1 - \alpha_\bfz \right)
	\\&\ge \frac{1}{e^2} \prod_{\bfz \in \Gamma_1}\left(1 - \alpha_\bfz \right)
	\\&\ge \frac{1}{e^2}
	\left(1 - \frac{1}{4\ell_1n } \right)^{n^2}.
\end{align*}
Finally, since 
$$ \left(1 - \frac{1}{4\ell_1n } \right)^{n^2} \approx \exp \left(-\frac{n}{4a\log(n)} \right), $$ it follows that 
$	\frac{1}{n} \log (Pr(\bfw \in \cD_n(L,t))) $ approaches 0 
as $ n \rightarrow \infty $. By plugging into (\ref{eq:da-rate}) the theorem statement holds.
\end{IEEEproof}

	\section{Reconstructing an Erroneous Multispectrum}\label{sec:rec-Erroneous}
	In this section, we address the problem of reconstructing strings from a multispectrum that suffered substitution errors. This family of multispectra is formally defined as follows.
\begin{definition}\label{def:ts-erroneous}
	Let $\bfw\in\Sigma^n$ be a string and $S_L(\bfw)$ is its $L$-multispectrum. A multiset $ U = \{\bfu_1,\dots, \bfu_{n-L+1}\} $ is called a \textbf{$(t,s)$-erroneous $L$-multispectrum of $\bfw$} if there exists a set of indices $ I_e(U) = \{{i_1},\dots,{i_m}\} \subset [n-L+1]$ where $ m \le t $ such that for every $ i \in [n-L+1] \setminus I_e(U) $, $ \bfu_i = \bfw_{i,L} $ and for every $ i \in I_e(U) $, $ d_H(\bfu_i, \bfw_{i,L}) \le s $. The \textbf{$(t,s)$-erroneous $L$-multispectrum ball of $\bfw$}, denoted by $\cB_{L,t,s}(\bfw)$, is defined to be the multiset
	$$\cB_{L,t,s}(\bfw) = \{ U \mid \textmd{$U$ is a $(t,s)$-erroneous $L$-multispectrum of $\bfw$} \}.$$
\end{definition}
We refer to the substrings of $ U $ at positions  $ I_e(U) $ as the \emph{erroneous substrings} of $ U $, and to the substrings at positions $ [n-L+1] \setminus I_e(U) $ as the \emph{correct substrings} of $ U $. 
\begin{example}\label{ex:ts-erroneous}	
	Let $ n=16, L=10 $ and $$ \bfx =  1011100010110111. $$ The $ L $-multispectrum of $ \bfx $ is 
	\begin{align*}
		\hspace{-1ex} S_L(\bfx) \hspace{-0.5ex} = \hspace{-0.5ex} \{&1011100010, 0111000101, 1110001011, 1100010110, \\&1000101101, 0001011011,0010110111 \}
	\end{align*}
	The multiset 
	\begin{align*}
	U = \{&0011100010, 0111000101, 1100001011, 1100011110, \\&1000101101, 0001011011, 0010110111 \}
	\end{align*}
	is a $ (3,1) $-erroneous $ L $-multispectrum of $ \bfx $. That is since by notating the substrings of $ U $ as $ \bfu_1, \dots, \bfu_7 $,  Definition~\ref{def:ts-erroneous} is satisfied for $ I_e = \{1,3,4\} $.
\end{example}

Let $ \bfw \in \Sigma^n $ be a string and $ U = \{\bfu_1,\dots, \bfu_{n-L+1}\} \in \cB_{L,t,s}(\bfw)$ be an erroneous spectrum. Note that if an entry of the input string $ \bfw $ can appear in $ U $ incorrectly more times than it appears correctly, we are not able to determine its correct value from $ U $. Hence, let $ \bfW_2(U) $ denote the maximum reconstructible-substring of $ U $, a string of length $ n $ that takes at every position $ i $ the majority value of the occurrences of  $ w_i $ in $ U $. Namely, for a multiset $ A $ over $ \Sigma $ we define the function $\text{maj}(A)$ that returns the element $ a \in A $ with the most appearances in $ A $. If there is more than one element of $ A $ that satisfies this requirement, the function $\text{maj}$ selects the first element in lexicographic order. For convenience, we sometimes apply the function $ \text{maj} $ on a vector instead of a multiset. Thus,
$$ \bfW_2(U) = (w'_1,\dots, w'_n) \text{ with } w'_j = \text{val}(U,j), 
$$ 
where
\begin{align*}
\hspace{-1ex}\text{val}&(U,j) \hspace{-0.5ex} =\hspace{-0.5ex} \text{maj}\{ (\bfu_i)_k \mid i \in [n-L+1], k \in [L], i\hspace{-0.2ex} +\hspace{-0.2ex} k \hspace{-0.2ex}-\hspace{-0.2ex} 1 \hspace{-0.2ex}=\hspace{-0.2ex} j \}.
\end{align*}

\begin{example}\label{def:ts-erroneous-cont}
	Following Example~\ref{ex:ts-erroneous}, $$ \bfW_2(U) =  0011100010110111, $$ which satisfies $ d_H(\bfW_2(U), \bfx) = 1 $. 
\end{example}

\begin{definition}\label{def:ts-erroneous-reconsructible}
	A string $\bfw$ is called an \textbf{\emph{$(L,t,s)$-reconstructible string}} if for any $(t,s)$-erroneous $L$-multispectrum $U \in \cB_{L,t,s}(\bfw)$, the maximal-reconstructible substring $\bfW_2(U)$ can be uniquely reconstructed from $ U $. 
\end{definition}

In order to have a controlled number of incorrect entries in $ \bfW_2(U) $, we add for the rest of this section the constraint $ t < L/2 $. This constraint ensures that for every $ U \in \cB_{L,t,s}(\bfw) $, all entries of $ \bfw $ besides the first and last $ 2t $ entries can not appear in $ U $ erroneously more times than their appear correctly. Therefore, $ \bfW_2(U) $ satisfies $\bfW_2(U)_{2t+1,n-4t}= \bfw_{2t+1,n-4t}$. Moreover, it follows that if for every $ j \in [2t] \cup [n-2t+1,n] $ the entry $ w_j $ equals val$(U,j)$, then $ \bfW_2(U) = \bfw $.

\begin{remark}
	Note that by altering the definitions of multispectrum and $ (L,d) $-substring distant strings to work with cyclic strings and cyclic multispectrum in which the substrings are read cyclically, the issues with the errors in the first and last $ 2t $ entries of the reconstructed string would have been solved. This will similarly resolve the same problems of incomplete multispectrum in Section~\ref{sec:rec-incomplete}. However, we choose to use the acyclic definitions, since they model better the DNA sequencing problem and are better consistent with previous works such as~\cite{GabMil18, EliGabMedYaa19}.
\end{remark}

In the next subsection we present a reconstruction algorithm for erroneous multispectra of $ (L,t,s)$-reconstructible strings, which is based upon the substring-distant property. Recall that a string $\bfw\in\Sigma^n$ is $(L,d)$-substring distant if the Hamming distance of its $L$-multispectrum is at least $d$. A natural question to ask is whether such strings exist and if so how many as a function of $n$, $L$, and $d$. We address this question in Section~\ref{subsec:IV_B} to provide conditions in which the redundancy of this set is at most one bit and when its asymptotic rate approaches 1. Then, in Section~\ref{subsec:IV_C}, we present encoding and decoding schemes for such strings which use a single bit of redundancy.


\subsection{Reconstruction Algorithm}
The main result of this subsection is summarized in the following theorem.
\begin{theorem}\label{th:naive}
	If a string $\bfx\in\Sigma^n$ is $ (L-1,4s+1) $-substring distant, then it is $(L,t,s)$-reconstructible for every $ t \in [n] $.
\end{theorem}
\begin{remark}
Notice that the constraint presented in Theorem~\ref{th:naive} is independent of $ t $, as the theorem holds for every $ t \in [n] $. This follows from the requirement of $ (4s+1) $-distance between the substrings, as seen in the upcoming proof. Alternative constraints, which are not independent of $ t $ and require smaller distance between the substrings, are presented in Section~\ref{sec:rec-Erroneous2}.
\end{remark}

The proof of Theorem~\ref{th:naive} is given by an explicit reconstruction algorithm, presented in Algorithm~\ref{alg:reconstruct-substitutions}. The algorithm receives an erroneous multispectrum $ U \in \cB_{L,t,s}(\bfx) $ for $ \bfx \in \SubDistSet_n(L-1,4s+1) $ and reconstructs the maximum reconstructible substring $\bfW_2(U)$. The algorithm uses the substring-distant property of $ \bfx $ to identify the correct order of the substrings of $ U $. Then, it takes for each entry of $ \bfx $ the majority vote of its occurrences in $ U $.

 \begin{algorithm}
	\caption{Reconstruct($ U,t,s $)}\label{alg:reconstruct-substitutions}
	\algorithmicrequire \ $U \in \cB_{L,t,s}(\bfx)$ for $ \bfx \in \SubDistSet_{n}(L-1,4s+1)$ \\
	\algorithmicensure \ $\bfW_2(U) $ the maximum reconstructible-substring of $U $
	\begin{algorithmic}[1]
		\State{Initialize $ B[1,\ldots,n] $ as an array of $ n $ empty vectors, set $ i = 1, A = U $}			
		\State{Pick $ \bfw_1 \in A $ such that for every other $ \bfw \in A $, $ d_H($Pref$_{L-1}(\bfw_1), $Suff$_{L-1}(\bfw) ) \ge 2s+1 $}\label{step:pickw1} 
		\State{Set $A = A\setminus \{\bfw_1\}$}
		\State{\textbf{For} every $ j = 1, \dots, L $, append $ (\bfw_1)_j $ to $ B[j] $ }		
		\While{$ |A| \neq 0 $}			
		\State{Pick $ \bfw_{i+1} \in A $ such that $ d_H( $Suff$_{L-1}(\bfw_{i}), $ Pref$_{L-1}(\bfw_{i+1})) \le 2s $} \label{step:match-suff-pref2} 
		\State{Set $A = A\setminus \{\bfw_{i+1}\}, i = i+1 $}
		\State{\textbf{For} every $ j = 1, \dots, L $, append $ (\bfw_i)_j $ to $ B[i+j-1] $ }		
		\EndWhile
		\State{Return $ \bfy = (y_1,\ldots,y_n)$ where $y_j = maj(B[j])  $}\label{step:makey} 
	\end{algorithmic}
	
\end{algorithm}

Let $ U =\{ \bfu_1, \dots, \bfu_{n-L+1} \} $ be the input set of the algorithm, ordered with respect to $ S_L(\bfx) $, similarly to Definition~\ref{def:ts-erroneous}. A demonstration of the execution of Algorithm~\ref{alg:reconstruct-substitutions} is presented in the next example. 
\begin{example}
	Let $ n,L,\bfx,U $ from Example~\ref{ex:ts-erroneous}. The string $ \bfx $ is $ (L-1,5) $-substring distant and therefore $ U $ is a valid input for Algorithm~\ref{alg:reconstruct-substitutions} with $ t=3, s=1 $. Let $ \bfu_1, \dots, \bfu_7 $ denote the elements of $ U $ similarly to Example~\ref{ex:ts-erroneous}. The algorithm picks at Step~\ref{step:pickw1} the substring $$ \bfw_1 = \bfu_1 = 0011100010 $$ since for every other $ i \in [2,7] $, 
	$$ d_H(001110001, \text{Suff}_{9}(\bfu_i)) \ge 3. $$
	Then, it continues to pick the other substrings of $ U $ in increasing order at Step~\ref{step:match-suff-pref2}, 
	since for every substring $ \bfu_i$ for $ i \in [1,6] $, only $ \bfu_{i+1} $ satisfies
	$$ d_H(\text{Suff}_9(\bfu_{i}),  \text{Pref}_9(\bfu_{i+1})) \le 2. $$
	For example, both $ \bfu_3,\bfu_4 $ are erroneous but yet satisfy
	$$ d_H(\text{Suff}_{9}(\bfu_3), \text{Pref}_{9}(\bfu_4))\hspace{-0.5ex} =\hspace{-0.5ex} d_H(100001011,110001111)\hspace{-0.5ex} =\hspace{-0.5ex} 2 $$ while for every $ i \neq 4 $, $$ d_H(\text{Suff}_{9}(\bfu_3), \text{Pref}_{9}(\bfu_i)) \ge 3. $$ Therefore, Algorithm~\ref{alg:reconstruct-substitutions}  holds for every $ j \in [n] $ all the occurrences of $ x_j $ in $ U $ inside the vector $ B[j] $. For example, $ B[1] = (0), B[5] = (1,1,0,1,1) $ and so on. Thus, following the construction of the result string in Step~\ref{step:makey}, the algorithm returns $$ \bfy = \bfW_2(U) = 0011100010110111. $$
\end{example}

We prove next the correctness of Algorithm~\ref{alg:reconstruct-substitutions}. 

\begin{lemma}\label{lem:ts-erroneous}
Given any $U \in \cB_{L,t,s}(\bfx)$ for $ \bfx \in \SubDistSet_{n}(L-1,4s+1)$, Algorithm~\ref{alg:reconstruct-substitutions} successfully returns $ \bfy =\bfW_2(U) $ at Step~\ref{step:makey}. 
\end{lemma}

\begin{IEEEproof}
	First, we show that the algorithm matches two substrings $ \bfw_i, \bfw_{i+1} $ in Step~\ref{step:match-suff-pref2} if and only if $ \bfw_i = \bfu_j $ and $ \bfw_{i+1} = \bfu_{j+1} $ for some consecutive $ \bfu_j, \bfu_{j+1} \in U $. That is, since from  $ U \in \cB_{L,t,s}(\bfx)$ we have that
	\begin{flalign*}
	 d_H(\text{Suff}&_{L-1}(\bfu_j), \text{Pref}_{L-1}(\bfu_{j+1})) 
	 	 \\&\le\begin{aligned}[t]
	&d_H(\text{Suff}_{L-1}(\bfu_j),\bfx_{j+1,L-1}) \\
	&+ d_H( \text{Pref}_{L-1}(\bfu_{j+1}),\bfx_{j+1,L-1}) 
	 \end{aligned}
	  \\& \le s+s = 2s,
	 \end{flalign*}
	and on the other hand, for $ \bfu_j, \bfu_{k} \in U $ with $ k \neq j+1 $, it follows from $ \bfx \in \SubDistSet_{n}(L-1,4s+1) $ and from $ U \in \cB_{L,t,s}(\bfx) $ that
	\begin{align*}
		 d_H(\text{Suff}&_{L-1}(\bfu_j),  \text{Pref}_{L-1}(\bfu_{k})) 
		 \\&\ge \begin{aligned}[t]
		  & d_H(\bfx_{j+1,L-1}, \bfx_{k,L-1}) - d_H(\text{Suff}_{L-1}(\bfu_j),\bfx_{j+1,L-1}) \\&-d_H(\text{Pref}_{L-1}(\bfu_{k}),\bfx_{k,L-1})
		  \end{aligned}
		  \\&\ge 4s+1 -s -s =  2s + 1.
	\end{align*}
	Using the same arguments, we pick at Step~\ref{step:pickw1} $ \bfw_1 = \bfu_1 $. Using a simple induction, it follows that for every $ i \in [n-L+1] $, $ \bfw_i = \bfu_i $. Hence, for every $ j \in [n] $, $$ B[j] = \{ (\bfu_i)_k \mid i \in [n-L+1], k \in [L], i + k - 1 = j \},  $$ and therefore the string  constructed by the algorithm in Step~\ref{step:makey} is $ \bfW_2(U) $.
\end{IEEEproof}

The proof of Lemma~\ref{lem:ts-erroneous} also verifies the correctness of Theorem~\ref{th:naive}. 

\subsection{Cardinality Analysis of $(L,d)$-Substring Distant Strings}\label{subsec:IV_B}
In this subsection we study the cardinality of the set of substring distant strings for different parameters of $L$ and $d$. For simplicity, all the results in the of this section are presented for the binary case. The next lemma assures that for given $d$ and for $n$ large enough, the redundancy of the set $\SubDistSet_n(L,d)$ is at most a single bit, when $L=2\log (n) +(d-1)\log (\log (n)) +\cO(1)$. 
\begin{lemma}\label{lem:Ld_red1}
For fixed $ d $, $L=2\log (n) +(d-1)\log (\log (n)) +\cO(1)$ and $n$ large enough, it holds that $\SubDistSize_n(L,d) \geq 2^{n-1}$ and hence the redundancy of the set $\SubDistSet_n(L,d)$ is at most a single bit. 
\end{lemma}
\begin{IEEEproof}
Let $L=2\log (n) +(d-1)\log (\log (n)) +C(d-1)$ for some positive constant $C$ that will be determined later. If a string is not an $(L,d)$-substring distant, then it contains at least two length-$L$ substrings which their Hamming distance is at most $d-1$. Notice that the size of the radius-$ (d-1) $ Hamming ball around any $ \bfv \in \Sigma^L $ is 
\begin{align*}
	|\cB_{d-1}(\bfv)| &= \sum_{i=0}^{d-1} \binom{L}{i} \le d\cdot \frac{L(L-1)\cdots (L-d+2)}{(d-1)!} \\&\le \frac{L^{d-1}}{(d-1)!/d} \le L^{d-1}+1.
\end{align*}
Hence, according to the union bound, the number of strings that are not $(L,d)$-substring distant can be bounded above by

\begin{align*}
n^2 2^{n-L} L^{d-1} & = 2^n  \frac{n^2(L^{d-1}+1)}{2^L} = 2^n  \frac{L^{d-1}+1}{(\log (n))^{d-1}2^{C(d-1)}} & \\
& \overset{(a)}{\leq} 2^n  \left(\frac{3\log (n)}{2^C\log (n)}\right)^{d-1} = 2^n  \left(\frac{3}{2^C}\right)^{d-1}, & 
\end{align*}
where inequality (a) holds for $n$ large enough. Hence, by choosing $C=\log(3)+1/(d-1)$ we get that the number of strings that are not $(L,d)$-substring distant is at most $2^{n-1}$, which accordingly implies that $\SubDistSize_n(L,d) \geq 2^{n-1}$.
\end{IEEEproof}

Our next result claims that the asymptotic rate of the set $\SubDistSet_n(L,d)$, when $L=\left\lceil a\log(n)\right\rceil$ and $a>1$, is 1. 
The proof follows the same outline as the proof presented in~\cite{EliGabMedYaa19IEEE} for $ L $-substring unique strings, and as the proof of Theorem~\ref{th:D}, and uses the asymmetric Lo\`asz local lemma as stated in Lemma~\ref{lem:lll}. We present the proof of this theorem here as well for the completeness of the results in the paper.
\begin{theorem}\label{th:E}
For fixed $d$, $a>1$, and $L=\left\lceil a\log (n) \right\rceil$, it holds that the asymptotic rate of the set $\SubDistSet_n(L,d)$ is $ 1 $. That is,
$$ \lim_{n\rightarrow\infty}\frac{\log(\SubDistSize_{n}(L,d))}{n} = 1.  $$
\end{theorem}
\begin{IEEEproof}
The size of $ \SubDistSet_n(L,d) $ will be estimated by a probabilistic approach. Assume that $ \bfw $ is a length-$ n $ string chosen uniformly at random over $ \Sigma $. In order to estimate the size $ \SubDistSize_n(L,d) $, we may estimate the probability $ Pr(\bfw \in \SubDistSet_n(L,d)) $ since $ \SubDistSize_n(L,d) = 2^n \cdot Pr(\bfw~\in~\SubDistSet_n(L,d))  $ and hence
\begin{equation}\label{eq:da-rate1}
	\begin{aligned}
		\lim_{n\rightarrow\infty}\frac{\log(\SubDistSize_{n}(L,d))}{n}
		= 1 + \limup{n}\frac{1}{n}\log(Pr(\bfw \in \SubDistSet_n(L,d))).
	\end{aligned}
\end{equation}

For integers $ i,j \in [n-L+1] $ we notate $ \bfz = (i,j) $ and denote by $ I_{\bfz} = \mathbbm{1}(\bfw_{i,L}\in \cB_{d-1}(\bfw_{j,L}))$ the indicator function of the event that the Hamming distance of the $ L $-substrings that start at positions $ i $ and $ j $ is less than $ d $. We denote the set of all admissible pairs,
\begin{align*}
	&\Gamma = \{ (i,j) \mid  i,j \in [n- L +1], i<j \}.
\end{align*}
Note that $ |\Gamma| \le n^2 $. We are interested in a lower bound on 
$$ Pr(\bfw \in \SubDistSet_n(L,d)) = Pr \left(\sum_{\bfz \in \Gamma} I_{\bfz} = 0\right). $$

Note that for every $ \bfz = (i, j) \in \Gamma $ it holds that $$Pr(I_\bfz) = \frac{|\cB_{d-1}(\bfw_{j,L})|}{2^L} \leq \frac{L^{d-1}+1}{2^L}. $$ Let $ \bfz = (i, j), \bfz' = (i', j') \in \Gamma  $. It is clear that if the substrings $ \bfw_{i,L}, \bfw_{j,L} $ do not overlap with $ \bfw_{i',L}$ or $ \bfw_{j',L} $, then the indicators $ I_{\bfz}, I_{\bfz'} $ are independent. We use Lemma~\ref{lem:lll} with a graph $ G = (V,E) $ such that $ V =  \Gamma $ and there is an edge $ \bfz \rightarrow \bfz' $  if at least one of $ \bfw_{i,L}, \bfw_{j,L} $ overlaps with $ \bfw_{i',L}$ or $ \bfw_{j',L} $. It follows that every $ \bfz \in  \Gamma $ has at most $ 4Ln $ neighbors in the graph.

We set the numbers $ \alpha_\bfz = \frac{1}{4Ln} $ for every $ \bfz \in \Gamma $. Then, it holds that
\begin{equation*}\label{eq:a11}
	\begin{aligned}
		\prod_{(\bfz, \bfz') \in E} (1-\alpha_{\bfz'}) &\ge  \left(1 - \frac{1}{4Ln } \right)^{4Ln } \ge \frac{1}{e}
	\end{aligned}
\end{equation*}
since the last expression approaches $ e^{-1} $ from above as {$ n~ \rightarrow~\infty $}. The condition of the lemma holds for every $ \bfz \in \Gamma $ since
\begin{align*}
	Pr(I_\bfz) &\le \frac{L^{d-1}+1}{2^{L}} \le\frac{L^{d-1}+1}{n^a} \overset{\text{(i)}}{\le} \frac{1}{4Ln} \cdot \frac{1}{e} \\&\le \alpha_\bfz \prod_{(\bfz, \bfz') \in E} (1-\alpha_{\bfz'}),	
\end{align*}
where  (i) follows from $ 4Ln = o(\frac{n^a}{L^{d-1}+1}) $ since $ L = \ceilenv{a \log(n)} $ and $ a > 1 $. 
By applying Lemma~\ref{lem:lll} we obtain
\begin{align*}
	Pr(\bfw \in \SubDistSize_n(L,d)) &\ge \prod_{\bfz \in \Gamma}\left(1 - \alpha_\bfz \right)
	\ge 
	\left(1 - \frac{1}{4Ln } \right)^{n^2}.
\end{align*}
Finally, since 
$$ \left(1 - \frac{1}{4Ln } \right)^{n^2} \approx \exp \left(-\frac{n}{4a\log(n)} \right), $$ it follows that 
$	\frac{1}{n} \log (Pr(\bfw \in \SubDistSize_n(L,d))) $ approaches 0 
as $ n \rightarrow \infty $. By plugging into (\ref{eq:da-rate1}) the theorem statement holds.
\end{IEEEproof}

\subsection{Encoding of $ (L,d) $-Substring Distant Strings}\label{subsec:IV_C}
In this section, a generic encoding algorithm is presented that uses a single redundancy bit in order to encode length-$ n $ strings that are $ (L,d) $-substring distant, for $$ L = 2 \log (n) + 2 (d-1+\epsilon) \log (\log (n)), $$
where $ \epsilon > 0 $ and for $ n $ large enough.
 Note that this value of $L$ is far from the value derived in Lemma~\ref{lem:Ld_red1} only by roughly $(d-1) \log (\log (n))$. 

First, we present some helpful definitions. 
Let $ \bfw,\bfw' \in \Sigma^n $ be strings such that $ d_H(\bfw, \bfw') \le \rho $ for an integer $ \rho \le n $. The construction $ EncDist_{n,\rho}(\bfw, \bfw') $ is taken from \cite{LevYaa18} and encodes the distance between $ \bfw, \bfw' $. Let $ p_1,\dots, p_{d_H(\bfw,\bfw')} $ denote the indices of the entries which $ \bfw, \bfw' $ do not agree upon. For every $ i \in [\rho] $ let $ \bfy_i \in \Sigma^{\log (n)}$ be the following value:
\[ \bfy_i = 
\begin{cases}
b(p_i) & i \le d_H(\bfw,\bfw') \\
0^{\log (n)} & \text{Otherwise}	
\end{cases}
\]
Thus, 	\[
EncDist_{n,\rho}(\bfw,\bfw') = \bfy_1 \circ \cdots \circ \bfy_{\rho}.
\]
Notice that the size of the output is independent of $ \bfw, \bfw' $ and equals $ \rho \cdot \log (n) $. We sometimes omit the parameter $ n $ if it is clear from the context. 

We utilize a marker substring, first introduced in~\cite{LevYaa18}, which we notate as a \emph{$ d$-auto cyclic string}. A string $ \bfu \in \Sigma^n $ is a $ d $-auto cyclic string, if it satisfies $$ d_H(\bfu, 0^i \circ \bfu_{1,n-i}) \ge d $$ for every $ 1 \le i \le d $. The authors of~\cite{LevYaa18} also presented a construction of such strings of length $ d \lceil \log (d) \rceil +2d $. Let $ \bfu_d $ denote a $ d $-auto cyclic string for the rest of this section.

Next, let $ \bfw \in \Sigma^k $ be a string for an integer $ k \le n $. Additionally, let $ r \le n $ be some integer. We want to construct a set of length-$ n $ strings, that contains all $ \bfy \in\Sigma^n $ that satisfy 
\begin{align}\label{eq:cb-t}
 d_H(\text{Pref}_{n}(\bfw \circ \bfy), \bfy) \le r.
\end{align}
 Therefore, we construct the \emph{concatenation ball of radius-$ r $} around $ \bfw $, denoted as $\cC\cB_{n,r}(\bfw)$. 
For this purpose, let $ m = \lceil n / k \rceil $ and let $ r_1, \dots, r_m $ be a series of integers such that $ \sum_{i=1}^m r_i \le r $. 
Furthermore, let $ \bfw_0, \bfw_1, \dots, \bfw_m $ be a series of substrings such that $ \bfw_0 = \bfw $ and for every $ i \in [m] $, $ \bfw_i \in \cB_{r_i}(\bfw_{i-1}) $. Thus, the string $ \text{Pref}_{n}(\bfw_1 \circ \cdots \circ \bfw_m) $  belongs to the set $ \cC\cB_{n,r}(\bfw)  $. Namely,
\begin{align*}
\cC\cB&_{n,r}(\bfw)  =
\\ &\left\{ \text{Pref}_{n}(\bfw_1 \circ \cdots \circ \bfw_m)\ \middle\vert 
\begin{array}{l}
\exists \{r_i \}_{i=1}^m \text{ s.t. } \sum_{i=1}^m r_i \le r,  \\
\bfw_0 = \bfw \text{ and } \forall i \in [m],\\  \bfw_i \in \cB_{r_i}(\bfw_{i-1})
\end{array}\right\}.
\end{align*}
One can verify that for every $ \bfy \in \Sigma^n$  that satisfies (\ref{eq:cb-t}), then $ \bfy  \in \cC\cB_{n,r}(\bfw) $. 

Algorithm~\ref{alg:ld-dist-encode1} receives a string $ \bfw \in \Sigma^{n-1} $, and outputs a string $ \bfx \in \SubDistSet_n(L,d) $. The algorithm shares ideas with the encoding scheme of  \emph{repeat-free words} from \cite{EliGabMedYaa19}, and consists of two main procedures, elimination and expansion. First, we append to $ \bfw $ a marker substring of length $ L/2+d+1 $ that contains the $ d $-auto cyclic string $ \bfu_d $, which is used by the decoder to identify the end of the information string. Then, at the elimination procedure we repeatedly look for substrings of length $ L $ that their Hamming distance is less than $ d $. When found, we remove the first of the substrings and encode the occurrence using the function $ EncDist_{L,d-1} $. Likewise, we eliminate occurrences of substrings of length $ L/2 $ that their Hamming distance from the $ (L/2) $-suffix of the string is less than $ d $. During this procedure, we ensure that the marker substring located at the suffix of the string remains intact. Later, at the expansion procedure we enlarge the string to length $ n $ by inserting substrings of length $ L/2 $ while making sure that the string remains $ (L,d) $-substring distant. We denote for the rest of this section $$ \ell =  L/2  = \log (n) + (d-1+\epsilon) \log (\log (n)). $$

\begin{algorithm*}
	\caption{LDEncode($ \bfw, L,d $)}\label{alg:ld-dist-encode1}
	\begin{algorithmic}[1]	
		\Require {A string $ \bfw \in \Sigma^{n-1} $}
		\Ensure {A string $\bfx \in \SubDistSet_n(L,d)$}
		\State Set $ \bfx = \bfw \circ  0 \circ 1^d \circ 0^{\ell-|\bfu_d|} \circ \bfu_d  $ \label{step:LDdtp-init}
		\Statex{\emph{Elimination}:}
		\While{exist indexes $ i < j $  such that $ d_H(\bfx_{i,L},\bfx_{j,L}) <  d $ \textbf{or} an index $ i \le |\bfx| -2\ell + |\bfu_d| $ where $ d_H(\bfx_{i,\ell},0^{\ell-|\bfu_d|}\circ \bfu_d) < d $ }\label{step:LDdtp-elim-while}
		\Case \textbf{1}: violating substrings $ \bfx_{i,L},\bfx_{j,L} $ exists				
		\If{$ i,j \in J_1 =  [|\bfx|-L -\ell -d, |\bfx|- L +1] $ ( $ \bfx_{i,L} $ intersects with the suffix $ 0 \circ 1^d \circ 0^{\ell-|\bfu_d|} \circ \bfu_d $)}
		\State{Remove $ \bfx_{i,L- (|\bfx|-L -\ell -d -i) + 1} $, append $  100 \circ b_{J_1}(i) \circ b_{J_1}(j) \circ EncDist_{L,d-1}(\bfx_{i,L}, \bfx_{j,L}) $ to the left of $ \bfx $ }\label{step:LDdtp-elim-if12-state}				
		\Else
		\State{Remove $ \bfx_{i,L} $, append $ \ 101 \circ b(i) \circ b(j) \circ 	EncDist_{L,d-1}(\bfx_{i,L}, \bfx_{j,L}) $ to the left of $ \bfx $}	\label{step:LDdtp-elim-if13-state}
		\EndIf
		\EndCase
		\Case \textbf{2}:{ a substring $ \bfx_{i,\ell} $ with $ i < |\bfx| -2\ell + |\bfu_d| $ such that $ d_H(\bfx_{i,\ell},0^{\ell-|\bfu_d|}\circ \bfu_d) < d $ exists }\label{step:LDdtp-elim-if2-cond}				
		\If{$ i \in J_2 =  [|\bfx|-2\ell -d, |\bfx|- 2\ell + |\bfu_d| -1] $ ( $ \bfx_{i,\ell} $ intersects with the suffix $ 0 \circ 1^d \circ 0^{\ell-|\bfu_d|} \circ \bfu_d $)}
		\State{Remove $ \bfx_{i,\ell- (|\bfx|-2\ell -d -i) + 1} $, append $ 11 \circ b_{J_2}(i) \circ EncDist_{\ell,d-1}(\bfx_{i,\ell}, 0^{\ell-|\bfu_d|}\circ \bfu_d) $ to the left of $ \bfx $ }	\label{step:LDdtp-elim-if22-state}			
		\Else
		\State{Remove $ \bfx_{i,\ell} $, append $  0 \circ b(i) \circ EncDist_{\ell,d-1}(\bfx_{i,\ell}, 0^{\ell-|\bfu_d|}\circ \bfu_d) $ to the left of $ \bfx $ }\label{step:LDdtp-elim-if23-state}
		\EndIf
		\EndCase	
		\EndWhile
		\State {\textbf{if } $|\bfx| \ge n $, return $ \bfx_{1,n}$}\label{step:after-elimination1}	
		\Statex{\emph{Expansion}:}
		\While{ $ |\bfx| < n $ } 
		\State{Set
			\[
			B = \bigg( \bigcup_{i \in [1,|\bfx|-\ell ]} \cB_{d-1}(\bfx_{i,\ell}) \bigg)  \cup \bigg( \bigcup_{i \in [|\bfx| - \ell + 1 , |\bfx|]} \cC\cB_{\ell,d-1}(\bfx_{i,|\bfx|-i+1}) \bigg)
			\]}\label{step:consB}	
		
		\State{ Pick $ \bfy \in \Sigma^{\ell} \setminus B $ and append $ \bfx = \bfx \circ \bfy $ }\label{step:LDdtp-exp-expand}	
		\EndWhile
		\State{Return $ \bfx_{1,n} $}\label{step:LDdtp-exp-ret}			
	\end{algorithmic}
\end{algorithm*} 

We prove the correctness of the algorithm in the next few claims. 
\begin{claim}\label{lem:LDdtp-elim-term}
	 Algorithm~\ref{alg:ld-dist-encode1} reaches Step~\ref{step:after-elimination1}, i.e. the elimination procedure terminates.
\end{claim}
\begin{IEEEproof}
	We prove by showing that at each iteration of the elimination loop, the length of $ \bfx $ decreases. We analyze each case of removal and insertion independently. All length comparisons are taken for a large enough $ n $.
	\\ \emph{Step~\ref{step:LDdtp-elim-if12-state}:} The minimal possible size of the removed substring at this step is achieved when $ i = |\bfx|-L $. Thus, the algorithm removes a substring of length at least $$ L - \ell -d +1  = \log (n) + (d-1+\epsilon) \log (\log (n))  -d + 1 $$ and inserts a smaller substring of length $$ (d-1) \log (L) + 2\log(\ell+d) + 3. $$
	\emph{Step~\ref{step:LDdtp-elim-if13-state}:} The algorithm removes a substring of length $$ L = 2\log (n) + 2(d-1+\epsilon)\log (\log (n)), $$ and inserts a substring of length $$ 2\log (n) + (d-1) \log (L) + 3 . $$ 
	\emph{Step~\ref{step:LDdtp-elim-if22-state}:} The minimal possible length of the removed substring at this step is reached when $ i = |\bfx|- 2\ell + |\bfu_d| -1 $. Therefore, the algorithm removes a substring of length at least
	\begin{align*}
		&\ell -  (|\bfu_d|+d+1) = \\&\log (n) + (d-1+\epsilon)\log (\log (n))  - (|\bfu_d|+d+1),
	\end{align*}  and inserts a smaller substring of length $$ (d-1) \log (\ell) + \log (|\bfu_d|+d+1) + 2. $$
	\emph{Step~\ref{step:LDdtp-elim-if23-state}:} The algorithm removes a substring of length $$ \ell = \log (n) + (d-1+\epsilon) \log (\log (n)),  $$ and inserts a substring of length
	\begin{align*}
	&\log(n) + (d-1)\log(\ell) + 1 =  \\&\log(n) + (d-1)\log\big(\log(n)\hspace*{-0.5ex}+\hspace*{-0.5ex}(d\hspace*{-0.5ex}-\hspace*{-0.5ex}1\hspace*{-0.5ex}+\hspace*{-0.5ex}\epsilon)\log(\log(n))\big) + 1,
	\end{align*}
	which is shorter for $ n $ large enough.
\end{IEEEproof}

\begin{claim}\label{lem:LDdtp-elim}
	At Step~\ref{step:after-elimination1} of Algorithm~\ref{alg:ld-dist-encode1}, the string $ \bfx $
	\begin{enumerate}[(1)]
		\item is $ (L,d) $-substring distant,
		\item ends with $ 0 \circ 1^d \circ 0^{\ell-|\bfu_d|} \circ \bfu_d $,
		\item contains no other $ \ell $-substring from $ \cB_{d-1}(\text{Suff}_\ell(\bfx)) $ besides its $ \ell $-suffix.
	\end{enumerate}
\end{claim}

\begin{IEEEproof}
Properties (1) and (2) follow immediately from the algorithm, since the loop continues as long as $ \bfx $ is not $ (L,d) $-distant, while ensuring that  Suff$_{\ell+d+1}(\bfx)$ is not touched.
	
	As for (3), from (2) we have that $ \text{Suff}_\ell(\bfx) = 0^{\ell-|\bfu_d|}\circ \bfu_d  $ and from the definition of a $ d $-auto cyclic string, for every $ i \in [d], $ $$ d_H(\bfu_d,0^i\circ (\bfu_d)_{1,|\bfu_d|-i-1}) \ge d $$  and hence for every $ i \in [|\bfx| - \ell - d, |\bfx| - \ell] $ we have  $$ d_H(\text{Suff}_\ell(\bfx), \bfx_{i,\ell}) \ge d. $$ For  $ i \in [|\bfx| - 2\ell + |\bfu_d|, |\bfx| - \ell -d - 1] $, the substring $ \bfx_{i,\ell} $ has $ 1^d $ starting at position $ (|\bfx| -\ell -d)-i $ while $ \text{Suff}_\ell(\bfx) $ has $ 0^d $ at this position. Other cases, for $ i <|\bfx| - 2\ell + |\bfu_d|  $, are eliminated at Step~\ref{step:LDdtp-elim-if2-cond}. 
\end{IEEEproof}

\begin{claim}\label{lem:LDdtp-exp-yexist}
	For every iteration of the expansion loop of Algorithm~\ref{alg:ld-dist-encode1}, the set $ \left|\Sigma^{\ell} \setminus B\right| $ constructed in Step~\ref{step:consB} is not empty.
\end{claim}
\begin{IEEEproof} 
	Using simple counting arguments we have that for every $ \bfw \in \Sigma^\ell $, the size of the radius-$(d-1)$ Hamming ball around $ \bfw $ satisfies $$ |\cB_{d-1}(\bfw)| = \sum_{d'=0}^{d-1} \binom{\ell}{d'} \le \ell^{d-1}. $$ Similarly, for every $ \bfw \in \Sigma^k $ with $ k \in [\ell] $, $ \cC\cB_{\ell,d-1}(\bfw) $ can be bounded using the same value. Thus,  
	the size of $B$ is bounded by $ n \cdot \ell^{d-1} $.		
	It is left to show that $  n \cdot \ell^{d-1} \le |\Sigma^\ell| =  2^\ell $. By taking a logarithm from both sides of the equation, we derive that it is necessary that 	
	\begin{align*}	
	\log (n) &+ (d-1)\log (\ell) \le \ell
	\end{align*}
	which is satisfied for $ n $ large enough by the value $$ \ell = \log (n) + (d-1+\epsilon)\log (\log (n)). $$ 	
\end{IEEEproof}

Let $ m $ denote the number of iterations of the expansion loop of Algorithm~\ref{alg:ld-dist-encode1} that were executed. For every $ k \in [m] $, let $ \bfx_k $ denote the value of $ \bfx $ at the end of the $ k $-th iteration, and let $ \bfy_k $ denote the string $ \bfy $ that the algorithm picked at Step~\ref{step:LDdtp-exp-expand} of that iteration. We notate by $ \bfx_0  $ the value of $ \bfx $ before the first iteration of the expansion loop.  In the next two lemmas, when referring to $ \bfx_k $ we sometimes omit the subscript $ k$ if it is clear from the context.

\begin{claim}\label{lem:LDdtp-exp-suffix}
	For every iteration $ k \in [m] $, the string $ \bfx = \bfx_{k-1} \circ \bfy_k  $ satisfies that for every $ i \in [1,|\bfx| - \ell],$ $$ d_H(\bfx_{i,\ell},\bfy_k) \ge d. $$ 		
\end{claim}
\begin{IEEEproof}
	According to the construction of $ B $, for every  $i \in [1,|\bfx_{k-1}| - \ell + 1]$ the ball  $\cB_{d-1}(\bfx_{i,\ell}) $ is contained in $ B $ and since $ \bfy_k \not \in B $ then $d_H(\bfx_{i,\ell},\bfy_k) \ge d$.
	Otherwise,  let $ i \in [|\bfx_{k-1}|-\ell+2, |\bfx_{k-1}|] $, assume in the contrary that $d_H(\bfx_{i,\ell},\bfy_k)  < d$ and thus $$ d_H(\text{Pref}_\ell(\bfx_{i,|\bfx|-i+1} \circ \bfy_k), \bfy_k) < d. $$ However, it follows that $ \bfy_k \in \cC\cB_{\ell,d-1}(\bfx_{i,|\bfx|-i+1}) $ which is a contradiction. Since $ |\bfx| -\ell = |\bfx_{k-1}| $, this concludes the proof.
\end{IEEEproof}

\begin{claim}\label{lem:LDdtp-exp-expand}
	For every iteration $ k \in [m] $, the string $ \bfx_{k} $ is $ (L,d) $-substring distant.
\end{claim}
\begin{IEEEproof}
	We prove the lemma by induction over the values of $ k $. 
	For the base case $ k = 1 $, let $ \bfx = \bfx_0 \circ \bfy_1 $ and assume in the contrary that there are two substrings $ \bfx_{i,L}, \bfx_{j,L} $ of Hamming distance less than $ d $. Since $ \bfx_0 $ is $ (L,d) $-substring distant from Claim~\ref{lem:LDdtp-elim} Statement (1), we only need to consider the cases where $ \bfx_{j,L} $ overlaps with $ \bfy_1 $. Therefore, using lengths considerations, $ \bfx_{j,L} $ contains  Suff$_{\ell}(\bfx_0) = 0^{\ell-|\bfu_d|}\circ \bfu_d $ at some position $ r \in [\ell] $. It follows that $ d_H(\bfx_{i+r,\ell}, \text{Suff}_{\ell}(\bfx_0)) < d $ which contradicts Claim~\ref{lem:LDdtp-elim} Statement (3).
	
	Next, we assume the lemma holds for $ \bfx_{k-1} $ with $ k \ge 1 $ and prove its correctness for $ \bfx = \bfx_{k-1} \circ \bfy_k $. Assume in the contrary that $ \bfx_{i,L}, \bfx_{j,L} $ satisfy  $ d_H( \bfx_{i,L}, \bfx_{j,L}) < d $. Using the induction assumption, we only need to consider the values of $ i,j $ where $ \bfx_{j,L} $ overlaps with $ \bfy_{k} $. Thus, it follows that $ \bfx_{j,L} $ contains the substring $ \bfy_{k-1} $, at some position $ r \in [\ell] $. However, this implies that $ \bfx_{i+r,\ell} $ is a substring of $ \bfx_{k} $ that satisfies $ d(\bfx_{i+r,\ell}, \bfy_k) < d $ while $ i + r \le |\bfx| - \ell $ which is a contradiction to Claim~\ref{lem:LDdtp-exp-suffix}.	
\end{IEEEproof}	

\begin{theorem}
	Algorithm~\ref{alg:ld-dist-encode1} successfully returns a string from $ \SubDistSet_n(L,d) $.
\end{theorem}
\begin{IEEEproof}
	If the condition in Step~\ref{step:after-elimination1} holds then according to Claim~\ref{lem:LDdtp-elim} Statement (1), $ \bfx $  is $ (L,d) $-substring distant. Since every substring of $ \bfx $  is $ (L,d) $-substring distant as well, the algorithm returns in this case a string that belongs to  $ \SubDistSet_n(L,d) $. Otherwise, from Claim~\ref{lem:LDdtp-exp-expand}, the algorithm returns a $ (L,d) $-substring distant string of length $ n $ at Step~\ref{step:LDdtp-exp-ret}.
\end{IEEEproof}


The decoding scheme receives $ \bfx $ which is an output of Algorithm~\ref{alg:ld-dist-encode1} and outputs $ \bfw \in \Sigma^{n-1} $. First, we look for the leftmost occurrence of the substring $ \bfv = 0 \circ 1^d \circ 0^{\ell-|\bfu_d|} \circ \bfu_d $ in $ \bfx $. According to Claim~\ref{lem:LDdtp-elim} Statement (2), the part of the string to the right of this substring was added during the expansion procedure and therefore we remove it from $ \bfx $. If the substring $ \bfv  $ is not present, we look for its longest prefix that is located as a suffix of $ \bfx $. The substring we found is a part of the substring $ \bfv  $ the algorithm added at Step~\ref{step:LDdtp-init} since the output of the algorithm is longer than the input. Thus, we can complete the substring to $\bfv  $ and receive $ \bfx $ at the stage of after the elimination procedure.

Next, we iteratively inverse the elimination procedure. We identify using the first three entries of $ \bfx $ the last step at which the data was encoded. If we encoded the data at Step~\ref{step:LDdtp-elim-if12-state} or Step~\ref{step:LDdtp-elim-if13-state}, we decode $ i,j $ from the function $ b $, and recover $ \bfx_{j,L} $ using $ \bfx_{i,L} $ and the encoded distance. If $ j \le i + \ell $ this has to be done carefully, by restoring every $ j-i $ entries of $ \bfx_{j,L} $ a time. If we encoded the data at Step~\ref{step:LDdtp-elim-if22-state} or Step~\ref{step:LDdtp-elim-if23-state}, we decode from the outputs of the functions $ b $ and $ EncDist_{L,d-1} $ the position $ i $ and the substring $ \bfx_{i,\ell} $, and insert the substring at position $ i $. We repeat this process until we obtain a substring of length $ n + \ell + d  $, and return its $ (n-1)$-prefix as  $ \bfw $.


	\section{Alternative Construction for Erroneous Multispectrum}\label{sec:rec-Erroneous2}
	One of the benefits of the construction in Section~\ref{sec:rec-Erroneous} is that it can tolerate a relatively large number of erroneous substrings. Namely, $t$ can have any value less than $L/2$. However, at the same time, the construction is identical for small and large values of $t$. In this section, we show an alternative construction for moderate values of $t$. More specifically, for $ t \le \log (n) / \log (\log (n)) $ we will have a weaker constraint than the one stated in Theorem~\ref{th:naive}. 
This will be given by strings that satisfy several constraints that are presented in Definition~\ref{def:reconstruction-constratints2}, along with a reconstruction algorithm, presented in Algorithm~\ref{alg:NLT1-EREC}. One difference between the algorithm presented in this section and the one from Section~\ref{sec:rec-Erroneous} is that the outcome here is not necessarily a length-$n$ string. However, the length of the output decoded string is at least $n-4t$ and it will be shown that it contains the substring $\bfx_{2t+1,n-4t}$ with no error. 

For the rest of this section, similarly to Section~\ref{sec:rec-incomplete}, we again use the integers $ \ell_1 = L - \floorenv{t/3} -1, \ell_2 =  L - \ceilenv{2t/3} -1, \ell_3 = L-t-1 $ and the sets $ I_2 = [n- \ell_2 - t+1,n- \ell_2 + 1], I_3 = [n- \ell_3 - t+1,n- \ell_3 + 1]$.

\begin{definition}\label{def:reconstruction-constratints2} 
	A string $\bfx\in\Sigma^n$ is said to satisfy the \textbf{$(n,L,t,s)$-erroneous reconstruction (EREC) constraints} if it fulfills the following three constraints. 
	\begin{enumerate}
		
		\item $ \bfx$ is a $ (\ell_1,2s+1)$-substring distant string.
		
		\item The first and last $ t +1 $ length-$ \ell_2$ substrings have a Hamming distance of at least $ 2s+1 $ from all other length-$ \ell_2$ substrings. Namely, for all $ i \in [t \plus 1], j \in [n- \ell_2 + 1]$ with $ i \neq j $ then	$ d_H(\bfx_{i,\ell_2}, \bfx_{j,\ell_2}) \ge 2s+1$ and for all $i \in [n-\ell_2+1], j \in I_2$ with $ i \neq j $, then $ d_H(\bfx_{i,\ell_2}, \bfx_{j,\ell_2}) \ge 2s+1$.
		
		\item The first $ t + 1 $ length-$\ell_3$ substrings have a Hamming distance of at least $ 2s+1 $ from the last $ t + 1 $ length-$\ell_3$ substrings. Namely, for all $ i \in [t \plus 1], j \in I_3 $, $d_H(\bfx_{i,\ell_3}, \bfx_{j,\ell_3}) \ge 2s+1$.
	\end{enumerate}
\end{definition}

Notice that these constraints are a variation of the constraints presented in Definition~\ref{def:reconstruction-constratints}, with the demand for inequality replaced with a demand of Hamming distance of at least $ 2s+1 $. In fact, when applying $ s=1 $ we receive that the $ (n,L,t,1) $-EREC constraints are exactly the $ (n,L,t) $-LREC constraints. 
For $n,L,t,s$, denote by $\cD_n(L,t,s)$ the set of all strings that satisfy the $(n,L,t,s)$-EREC constraint and let $D_n(L,t,s) = |\cD_n(L,t,s)|$.

Let $ U \in \cB_{L,t,s}(\bfx) $ for $ \bfx \in \Sigma^n $. Next, we modify the definition of the maximal-reconstructible string of $ U $, and denote it as $ \bfW_3(U) $. Let $ V = \{\bfu_{i_1},\dots, \bfu_{i_m} \} \subseteq U $ denote for the rest of this section a subset of $ U $ where $ m \ge n-t $. Let $ I(V) = [i_1,i_m + L-1] $ be the set of positions of $ \bfx $ entries that appear in $ V $ (the segment is continuous since $ t < L $), and let $ n(V) = i_m + L -i_1 $ denote its size. We define for every $ j \in I(V) $, the function cons$(V,j)$ which returns the consented value of the entry $ x_j $ in $ V $, or an error if such a value does not exist. That is,
\[
\text{cons}(V,j) = \begin{cases}
a & \text{if for every } k \in [m] \text{, where } \\&j \in [i_k, i_k + L - 1], (\bfu_{i_k})_{j - i_k + 1} = a
\\	error & \text{otherwise}
\end{cases}
.\]
We say that $ V $ has a \emph{consensus} if for every  $ j \in I(V) $, cons$ (V,j) \neq error $. For every such $ V $ we define its consented string to be 
$$ \text{cons}(V) = (w_1,\dots,w_{n(V)}) \text{ with } w_i = cons(V,i+i_1-1). $$
Finally, Let $ V^* \subseteq U $ denote the largest subset of $ U $ that has a consensus. Thus, we define $ \bfW_3(U) = \text{cons}(V^*). $
For simplicity, we say that $ V^* $ is the subset of $ U $ that contains only its correct substrings. Namely,  $$ V^* = \{\bfu_i \in U \mid i \not \in I_e(U) \}. $$ In exceptional cases $ V^* $ might be a different set and we will analyze those cases later in this section. 

We modify the definition of $(L,t,s)$-reconstructible strings in Definition~\ref{def:ts-erroneous-reconsructible} to use $ \bfW_3(U) $ instead of $ \bfW_2(U)$. 
The main theorem of this subsection is presented next, in Theorem~\ref{th:EREC reconstruction}. The proof of this theorem is given by an explicit reconstruction algorithm.
\begin{theorem}\label{th:EREC reconstruction}
	Every string $\bfx \in \cD_n(L,t,s)$ is an $(L,t,s)$-reconstructible string.
\end{theorem}

Algorithm~\ref{alg:NLT1-EREC}  uses the stitching and reconstruction algorithms presented in Section~\ref{sec:rec-incomplete}. First, we extend Reconstruct$ (U,t) $ to notice errors by appending an additional step at its end. If Reconstruct$ (U,t) $ reaches this step, it returns $ error $. It follows that such a result occurs if Reconstruct$ (U,t) $ fails to successfully reconstruct at Steps 2, 3 and 4. 
Algorithm~\ref{alg:NLT1-EREC} receives $ U \in \cB_{L,t,s}(\bfx) $ for $ \bfx \in \cD_n(L,t,s) $ and reconstructs $ \bfW_3(U) $. First, it invokes the stitching algorithm with $ U $ and $ t'=0 $ and receives a set of substrings denoted by $ A_0 = \{\bfy_1,\dots, \bfy_r \} $. The fact that $ \bfx $ is $ (L-\floorenv{t/3}-1,2s+1)$-substring distant (and in particular $ (L-1,2s+1)$-substring distant) ensures that errors are present only at the edges of the substrings of $ \bfy_1,\dots,\bfy_r $. Based on this observation, we look for a  subset of $ U $ with maximal size that has a consensus. For every candidate spectrum $ V $, we use Reconstruct$ (V,t) $ to identify if $ V $ has a consensus, and return the consented string. 

Before presenting the  algorithm, we define for a string $ \bfw \in \Sigma^n $ and an integer $ k \in [n] $, its \emph{striping ball} of radius $ k $, denoted by $ SB_{k}(\bfw) $, as the set 
\[
SB_{k}(\bfw) = \{ \bfw_{i+1, n-j-i+1} \mid 0 \le i+j \le k \} .
\]



\begin{algorithm}[H]
	\caption{$(n,L,t,s)$-EREC Reconstruction}\label{alg:NLT1-EREC}
	\begin{algorithmic}[1]
		\Require  $  U \in \cB_{L,t,s}(\bfx) $ for $ \bfx \in \cD_n(L,t,s) $ 
		\Ensure $ \bfW_3(U) $ the maximum reconstructible-substring of $U $ 
		\State{$ A_0 = \text{Stitch}(U,0) $, denote $ A_0 =\{\bfy_1,\dots, \bfy_{r}\}  $}\label{step:init-stitch}
		\State{Set $ \alpha = t-\lceil(r-1)/2\rceil+1 $, and construct $$ B = \{ \bigcup_{j=1}^r S_L(\bfw_j) \mid \forall_{j \in [r]} : \bfw_j \in SB_{\alpha}(\bfy_j) \},  $$ $$ \forall {\rho \in [0,t]} \text{ set } B_\rho = \{V \in B \mid |V| = |U| - \rho\}. $$ }\label{step:init-setB}
		\For{every $ \rho 
			\in [0,t] $}	\label{step:nlt1erec-attempt}	
		\If{exists $ V \in B_\rho $ with Reconstruct$(V,t) \neq error $}
		\State{return $  \text{Reconstruct}(V,t)  $ . }\label{step:return-result}
		\EndIf
		\EndFor		
	\end{algorithmic}	
\end{algorithm}

We prove the correctness of Algorithm~\ref{alg:NLT1-EREC} in the next few claims. Let $ \bfu_1, \dots, \bfu_{n-L+1}  $ denote the elements of $ U $ with respect to the order of $ S_L(\bfx) $.
\begin{claim}\label{lem:stitch-correct}
	At Step~\ref{step:init-stitch}, in $\text{Stitch}(U,0)$ we stitch two substrings $ \bfw, \bfw' $ if $ \text{Suff}_L(\bfw) = \bfu_i $ and $ \text{Pref}_L(\bfw') = \bfu_{i+1} $ for some $ \bfu_i, \bfu_{i+1} \in U $. 
\end{claim}
\begin{IEEEproof}
	Let  $ \bfw, \bfw' $ denote two substrings with $ \text{Suff}_L(\bfw) = \bfu_i $ and $ \text{Pref}_L(\bfw') = \bfu_{j}$ that were stitched in $\text{Stitch}(U,0)$ using $ \text{Suff}_{L-1}(\bfw) = \text{Pref}_{L-1}(\bfw') $. Assume in the contrary that $ j \neq i+1 $. From $ (n,L,t,s) $-EREC Constraint 1 it follows that $ \bfx $ is $ (L-1, 2s+1) $-substring distant. Thus, since $ U \in \cB_{L,t,s}(\bfx) $ we have that 
	\begin{align*}
	d_H(\text{Suff}&_{L-1}(\bfu_i),  \text{Pref}_{L-1}(\bfu_{j})) 
	\\&\ge \begin{aligned}[t]
	& d_H(\bfx_{i+1,L-1}, \bfx_{j,L-1}) - d_H(\text{Suff}_{L-1}(\bfu_i),\bfx_{i+1,L-1}) \\&-d_H(\text{Pref}_{L-1}(\bfu_{j}),\bfx_{j,L-1})
	\end{aligned}
	\\&\ge 2s+1 -s -s =  1.
	\end{align*}
	which is a contradiction. 	
\end{IEEEproof}

For every substring $ \bfy_j \in A_0 $, we notate by $ Y_j $ the \emph{substring-set} of $ \bfy_j $ which is the set that contains all the substrings of $ U$  that are substrings of $ \bfy_j $. That is, $ Y_j =S_L(\bfy_j) = \{\bfu_{i_j},\ldots,\bfu_{i_j + m_j}\} $ where $ m_j = |\bfy_j| - L $. We say that $ Y_j $ is erroneous if it contains an erroneous substring.
\begin{claim}\label{lem:after-init-stitch}
	After Step~\ref{step:init-stitch}, the following conditions hold. 
	\begin{enumerate}[(1)]		
		\item There are at least $\lceil(r-1)/2\rceil $ substrings in $ A_0 $ that their substring-set contains an erroneous substring,
		\item $ r \le 2t+1 $,
		\item For every $ j \in [r] $, the substring-set $ Y_j $ contains at most $ \alpha = t-\lceil(r-1)/2\rceil+1 $ erroneous substrings. 
	\end{enumerate} 
\end{claim}
\begin{IEEEproof}
	Assume that $ \bfy_1,\dots, \bfy_r $ are ordered in correspondence to the positions of their substring-sets; that is, $ i_1 = 1, i_r+m_r = n-L+1 $ and for every $ j \in [2,r] $, $ i_{j-1}+m_{j-1} + 1 = i_j  $. Thus, using Claim~\ref{lem:stitch-correct}, for every $ j \in [1,r-1] $, at least one of $ \bfu_{i_{j}+m_{j}}, \bfu_{i_{j+1}} $ is erroneous. Therefore, the minimal number of erroneous substring-sets is $ \lceil(r-1)/2\rceil $. It can be   achieved for example when only the substring-sets with even indecies are erroneous (and additionally $ Y_r $ if $ r $ is odd), thus proving statement (1).
	Statements (2) and (3) follows immediately from (1) and from the fact that there are at most $ t $ erroneous substrings, and thus at most $ t $ erroneous substring-sets. 
\end{IEEEproof}

\begin{claim}\label{lem:where-errors}
For every $ \bfy_j \in A_0 $ there exists a substring $ \bfw_j \in SB_\alpha(\bfy_j) $ such that $ S_L(\bfw_j) $ contains only the correct substrings of $ Y_j $. 
\end{claim}
\begin{IEEEproof}
	First, if $ Y_j $ contains no erroneous substrings, picking $ \bfw_j = \bfy_j $ yields the required result. Else, let $ g,h $ denote the positions of the erroneous entries of $ \bfy_j $ that are closest to the center of the substring, from the left and right receptively. Namely, denote by $ I_e(\bfy_j) $ the positions of the erroneous entries of $ \bfy_j $, so we get that 
	$$ g = \max\left\{i \in I_e(\bfy_j) \mid i \le \floorenv{\frac{|\bfy_j|}{2}}  \right\},$$ 
	$$ h = \min\left\{i \in I_e(\bfy_j) \mid i \ge \ceilenv{\frac{|\bfy_j|}{2} } \right\}. $$
	Assume without loss of generality that both $ g,h $ exist. From Claim~\ref{lem:stitch-correct}, all the substrings of $ Y_j $ that contain those entries agree with the errors. Therefore, the sets $ E(g) = Y_j \cap \{\bfu_{g-L+1}, \dots ,\bfu_{g} \}, E(H) =  Y_j \cap \{\bfu_{h-L+1}, \dots, \bfu_{h} \}  $ contain the erroneous entries $ (\bfy_j)_g ,(\bfy_j)_h $, respectively. Since from Claim~\ref{lem:after-init-stitch} statement (3), $ Y_j $ contains at most $ \alpha $ erroneous substrings and since $ k < L $, we have $$ E(g) = \{\bfu_{i_j}, \dots, \bfu_{g} \}, E(h) = \{\bfu_{h-L+1},\dots,\bfu_{i_j + m_j}\} $$ and hence $ |E(g)| + |E(h)| \le \alpha  $. Finally, we can pick $ \bfw_j = (w_{g+1}, \dots, w_{h-L}) $ which satisfies $ |\bfw_j|\le |\bfy_j| - k $ or alternatively $ \bfw_j \in SB_\alpha(\bfy_j) $. It is ensured from the selection of $ g,h $ that $ \bfw_j $ is error-free. 
\end{IEEEproof}

Let $ B^* $ denote the union $ \bigcup_{\rho=0}^t B_\rho $. Notice that every spectrum $ V \in B^* $ satisfies $ V \subseteq U $ and $ |V| \ge |U| - t $. 

\begin{claim}\label{lem:recons-agree}
	For every $ V \in B^* $, Reconstruct$ (V,t) $ returns $ error $ if and only if there exist $ i \in I(V)  $ such that cons$(V,i) = error $. 
\end{claim}
\begin{IEEEproof} 
	In the proof of the reconstruction algorithm in Lemma~\ref{lem:alg-reconstruct-incomplete} in Section~\ref{sec:rec-incomplete}, we use arguments that rely on an $ (n,L,t) $-LREC Constraint to justify that we stitched at some iteration $ k $ of the stitching algorithm two substrings $ \bfw, \bfw' $ if $ \text{Suff}_L(\bfw) = \bfu_i $ and $ \text{Pref}_L(\bfw') = \bfu_{i+k} $. We can replace those arguments with arguments that are based on $ (n,L,t,s) $-Constraint like we used in the proof of Claim~\ref{lem:stitch-correct}, to prove the same claim for the operation of Reconstruct$ (V,t) $. 
	
	Thus, we can derive that Reconstruct($ V,t $) returned a successful result if and only if all the substrings of $ V $ were stitched in the correct order. Following previous observation, it is only possible if for every $ i \in I(V) $, cons$(V,i)  \neq error $ and in this case the result string is $ \text{cons}(V) $.
\end{IEEEproof}

\begin{claim}\label{lem:found-V}
	The set $ V^* = \{\bfu_i \in U \mid i \not \in I_e(U) \} $ belongs to $ B^* $ and Reconstruct$ (V^*,t) $ is successful. 
\end{claim}
\begin{IEEEproof} 
	Following Claim~\ref{lem:where-errors}, there exists for every $ \bfy_j \in A_0 $ a substring $ \bfw_j \in SB_\alpha(\bfy_j) $ such that $ S_L(\bfw_j) $ contains only the correct substrings of $ Y_j $. By picking such $ \bfw_j $ for all substrings of $ A_0 $ that contained errors, we construct the set $ V^* $. From the definition of $ B $ in Step~\ref{step:init-setB}, $  V^* \in B $. Moreover, since there are at most $ t $ erroneous substrings, $  V^* \in B^* $ as well. Since $  V^* $ contains no erroneous entries, it satisfies the conditions of Claim~\ref{lem:recons-agree} and thus Reconstruct$( V^*,t)$ is successful. 
\end{IEEEproof}

\begin{claim}\label{lem:size-B}
	The size of the set $ B^* $ satisfies $ |B^*| \le n $
\end{claim}
\begin{IEEEproof}
	For every $ \bfy_j \in A_0 $, the size of $  SB_{\alpha}(\bfy_j) $ can be a viewed as the number of possible selections of $ n_{j,1}, n_{j,2} $ entries to remove from the left, right of $ \bfy_j $ respectively, with $ n_{j,1}+ n_{j,2}\le \alpha $. This reflects removing $ n_{j,1}+ n_{j,2} $ substrings from $ Y_j $, which are also absent from a set $ V \in B $. 
	Therefore, for every $ p \le t $ we can bound the size of $ B_p $ with the number of possible solutions to the equation $$ \sum_{j=1}^{r} (n_{j,1}+ n_{j,2}) = p, $$ where for every $ j \in [r] $, $n_{j,1}+ n_{j,2} \le \alpha$. This value can be bounded using $$ |B_p| \le \binom{p+2r-1}{p}, $$ and therefore $$ |B^*| \le \sum_{p=0}^{t} |B_p| \le \sum_{p=0}^{t} \binom{5t+1}{p} \le (5t+2)^t. $$ Since $ t \le \log (n) / \log ( \log (n)) $, for a large enough $ n $ we have that $ |B^*| \le n $.
\end{IEEEproof}	

Lastly, we conclude with the following lemma.
\begin{lemma}\label{lem:algNLT1-EREC}
	Algorithm~\ref{alg:NLT1-EREC} returns $ \bfW_3(U)$ and has a polynomial-time complexity.
\end{lemma}
\begin{IEEEproof}
If follows from Claim~\ref{lem:found-V} that the algorithm finds $ V^* \in B_\rho $ for $ \rho = |I_e(U)| $. Thus, the algorithm returns $ \text{cons}( V^*) = \bfW_3(U) $ at step~\ref{step:return-result}. 

The time complexity of the algorithm is dominated by the loop of Step~\ref{step:nlt1erec-attempt}. Since for any set $ V \subseteq U $ the time complexity of Reconstruct $(V,t)  $ is a polynomial of $ n $, and from Claim~\ref{lem:size-B} the size of $ B^* $ is  a polynomial of $ n $, Algorithm~\ref{alg:NLT1-EREC} has polynomial-time complexity. 
\end{IEEEproof}
The proof of Lemma~\ref{lem:algNLT1-EREC} also completes the proof of Theorem~\ref{th:EREC reconstruction}.

Finally, we analyze the results of Algorithm~\ref{alg:NLT1-EREC} in cases where $ V^* \neq \{\bfu_i \in U \mid i \not \in I_e(U) \} $.
An erroneous substring can belong to  $ V^* $ when the majority value of an entry is not the correct entry. This can occur if an entry appears erroneously in the majority of their occurrences in $ U $, or in a subset of $ U $ after removal of some incorrect substrings. By addressing to the majority values of  entries rather than their correct values in the proof of Algorithm~\ref{alg:NLT1-EREC}, we can derive that the algorithm constructs $ V^* \in B^* $ in Step~\ref{step:init-setB} and thus reconstructs $ \bfW_3(U) $ in this case as well. 	
Since $ t < L/2 $ and $ |V^*| \ge |U|-t $, $ |\bfW_3(U)| \ge n-t $. Furthermore, since there are at most $ t $ erroneous substrings in $ V^* $, for every $ i \in [2t+1,n-2t] $, $ \text{cons}(V^*,i) = \bfx_i $. Thus, $$ \bfW_3(U)_{i_1+2t+1, n-4t} = \bfx_{2t+1,n-4t}, $$
where $ i_1 $ is the position of the first entry of $ \bfx $ that appears in $ V^* $.


	\section{Conclusion}\label{sec:concl}
	This paper studied the reconstruction of strings based upon noisy versions of their multispectrum. In the first model, we assumed that not all substrings in the multispectrum are read and in the second, it was assumed that all substrings are read, however several of them can be erroneous. In each case we studied code constructions of strings that can be uniquely reconstructed from the noisy version of the multispectrum. The cardinalities of the codes are studied along with specific code constructions. An important ingredient in our constructions is the set of $(L,d)$-substring distant strings. We studied when the redundancy of this set is at most a single bit and when its asymptotic rate approaches 1. We also presented specific encoding and decoding maps for this constraint. While this work studied only the binary case, most of the results in the paper can be extended for the non-binary case as well. While the results in the paper provide a significant contribution in the area of coding for reconstruction from substrings spectrum, there are still several interesting problems which are left open. Some of them are the constructions and analysis of more cases for the values of $L$, $t$, and $d$, and especially studying the erroneous case of edit errors together with losses of substrings.
	
	\section*{Acknowledgement}
	The authors would like to thank the associate editor Amin Gohari and the three anonymous reviewers for their very helpful comments. They are also thankful to Ohad Elishco for helpful discussions on the proofs of Theorem~\ref{th:D} and Theorem~\ref{th:E}.

\bibliographystyle{IEEEtranS}
\bibliography{mybib}

\end{document}